\documentclass{emulateapj}  % [preprint1]
\usepackage{lineno}
\usepackage{color}

\newcommand{\msun} {${\cal M}_\odot$}

\begin{document}

\title{Statistics of wide pre-main sequence binaries in the Orion OB1 association}

\author{Andrei Tokovinin}
\affiliation{Cerro Tololo Inter-American Observatory | NSF's NOIRlab, Casilla 603, La Serena, Chile}
 \author{Monika G. Petr-Gotzens}
\affiliation{   European          Southern         Observatory,
  Karl-Schwarzschild-Strasse,   2  D-85748  Garching   bei  M\"unchen,
  Germany}
\affiliation{Universit\"ats-Sternwarte, Ludwig-Maximilians-Universit\"at M\"unchen, Scheinerstr 1, D-81679 M\"unchen, Germany} 
\author{Cesar Brice\~no}
\affiliation{Cerro Tololo Inter-American Observatory, | NSF's NOIRLab, Casilla 603, La Serena, Chile}
%\affiliation{Cerro Tololo Inter-American Observatory, Casilla 603, La Serena, Chile}

\email{atokovinin@ctio.noao.edu}
%\correspondingauthor{Andrei Tokovinin}

\begin{abstract}
Statistics of low-mass pre-main sequence binaries  in the Orion OB1 association with separations ranging from 0\farcs6 to  20\arcsec ~(220 to 7400 au at 370 pc)  are studied  using images  from  the VISTA  Orion mini-survey  and astrometry  from   Gaia.   The  input sample  based  on the  CVSO catalog contains  1137 stars of  K and M  spectral types (masses between 0.3 and 0.9 \msun), 1021 of which are considered to be association members. There  are 135  physical binary  companions to  these stars  with mass ratios   above  $\sim$0.13.    The  average   companion   fraction  is 0.09$\pm$0.01 over  1.2 decades in  separation, slightly less  than,  but still consistent with, the field.  We found a difference between the Ori OB1a and OB1b groups, the latter being richer in  binaries by a factor 1.6$\pm$0.3.  No overall dependence  of the  wide-binary frequency  on the  observed underlying stellar density  is found, although  in the Ori OB1a  off-cloud population these binaries  seem to avoid  dense clusters.  The  multiplicity rates  in Ori  OB1 and  in sparse  regions like Taurus differ  significantly, hinting that  binaries in the  field may originate from a mixture of diverse populations. 
\end{abstract}

%\maketitle

%---------------------------------------------------------
\section{Introduction}
\label{sec:intro}

Orbital parameters  and mass  ratios of binary  stars depend  on their formation environment. It is  known that star formation regions (SFRs) of low  stellar density, like Taurus-Auriga,  spawn a rich  binary population, including a substantial number of very wide pairs \citep{Joncour2017}. In  contrast,  in  more  dense  SFRs the  binary  fraction  is  lower, comparable       to       the       field      binary       population \citep{Duchene2013,King2012}. It is generally accepted that most stars in the  field were  formed in relatively  dense environments  and that some young wide binaries  were destroyed by dynamical interaction with neighboring  stars. However,  \citet{Duchene2018} found  an  excess of close (10--60  au) binaries  in the dense  Orion Nebula Cluster (ONC), compared  to the field.   These  close binaries  are   not susceptible to dynamical  disruption  \citep{Parker2014}.  Critical examination  of binary statistics in several  nearby SFRs has led  \citet{Duchene2018}  to  the disconcerting conclusion  that none  of those groups is compatible with the binary statistics in the field.   However, the excess of binaries with  separations $<$60 au in the ONC has later been contested by \citet{defurio2019}.

The ongoing  debate on the origin  of the field  binary population and the  role of  SFR density  and dynamical  interactions in  shaping the binary separation  distribution  stimulates  further  observational  studies. Currently available data on  multiplicity statistics suffer from large errors owing to  the small size of available  samples and from various biases  caused  by   observational  constraints  or  sample  selection effects. Modern large--scale surveys  and catalogs change the landscape by providing large  and homogeneous data sets.  For  example, the   Gaia  census  of nearby  wide  binaries  gave  new insights  on  the distribution     of    their     separations    and     mass    ratios \citep{El-Badry2019}.   A  sample  of  $\sim$600 stars  in  the  Upper Scorpius SFR has been recently observed with high angular resolution to refine the binary statistics \citep{TokBri2019}. 

\begin{figure}
	\epsscale{1.1}
	\plotone{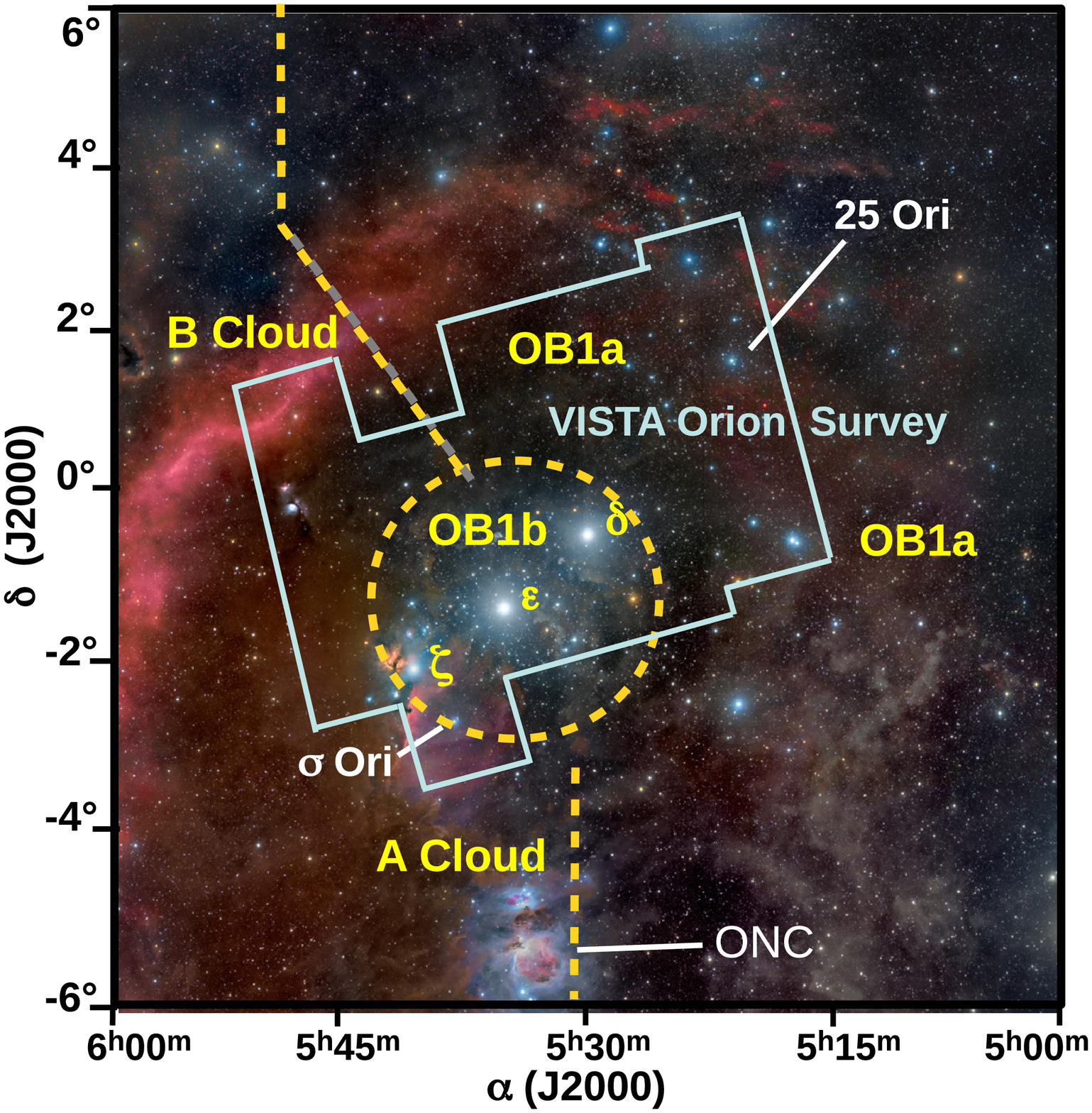}
\caption{Wide field optical image of the area encompassed by the CVSO in the Orion OB1 association, showing as an irregular polygon the approximate footprint of the VISTA Orion Survey \citep{Petr2011}. The Orion OB1b sub-association is the region within the dashed line circle, as in \cite{briceno2005}. The OB1a sub-association is the area to the west of the OB1b region and the dashed lines north and south of it, which roughly mark the limits of the Orion A and B molecular clouds, as indicated by the labels. We also indicate the location of the three Orion belt stars $\delta$, $\epsilon$, and $\zeta$ Ori, as well as the $\sigma$ Ori and 25 Ori clusters.  The Orion Nebula Cluster (ONC) is also indicated. Photo courtesy of Rogelio Bernal Andreo (http://www.deepskycolors.com/).
			\label{fig:cvso_vista} 		}
	\end{figure}

Here we use  the opportunity to learn about  young binaries offered by the combination of three modern  surveys: CVSO, VISTA Orion, and  Gaia. The CVSO \citep[CIDA Variability Survey of Orion;][]{CVSO} was an optical, multi-epoch imaging survey that produced a  large  sample of  pre-main sequence (PMS) stars   across $\rm \sim 180 \deg^2$ in the Orion OB1 association, spanning all the region between $\alpha_{J2000}= 5^{|rm h} - 6^{\rm h}$,  and $\delta_{J2000}=-6\degr$ to +6\degr ~(Figure \ref{fig:cvso_vista}), with an average resolution of 3\arcsec ~(equivalent to 1112 au at 370 pc) as measured in the original CVSO images. The young stars were selected based on  photometric variability  and confirmed by follow-up spectroscopy. These PMS stars  are mostly located outside the Orion A and B  molecular clouds \citep{maddalena86}  and we refer to them as off-cloud PMS stars. The off-cloud stars  have on average low  extinction ($A_V \la 0.5$ mag). The spectral types  range, mostly, from M5  to K0 and  correspond to masses   $\sim 0.3 - 0.9$  \msun.   The off-cloud CVSO PMS stars are localized in the two main sub-associations  in which Ori OB1 was been traditionally subdivided, namely Ori OB1a and Ori OB1b \citep{blaauw64,wh77}, with further sub--clustering within each group \citep{CVSO}. 

The  VISTA  Orion  survey \citep{Petr2011}   covers a 30 square degree  area toward the Orion belt  and  was designed to overlap in  large  parts  with  the  CVSO footprint, although the total area covered is much smaller, as shown in Figure \ref{fig:cvso_vista}. Its  near infra-red  (nIR)  images  in the  $Z, Y, J, H, K_s$ photometric bands  have a typical  stellar point spread function FWHM  (full width at half maximum) resolution of 0\farcs9, and allow us to  detect binaries  down to  0\farcs6 separation     (projected separation of $\sim$220 au  at 370\,pc distance).   Statistics  of  wide   binaries  can  be  studied  after accounting  for  chance  pairs   of  unrelated  field  stars  (optical companions).  However, distinguishing statistically true binaries from random asterisms  becomes  progressively  uncertain  with  increasing separation and magnitude difference.

The   Gaia Data Release 2 \citep{Gaia},  hereafter GDR2, contains parallaxes and proper  motions (PMs) of most bright stars  in the VISTA Orion catalog, allowing  a much more  reliable distinction between  real and optical  pairs.  At  the same  time, it  helps to  clean the  main CVSO sample. However,  GDR2 has  its own problems,  mostly caused  by close (unresolved)  binaries.   As a  result,  reliable  GDR2 astrometry  is available  for most, but  not all,  stars and  companions in  Ori OB1. This difficulty  can be partially  circumvented by using the  CVSO and VISTA Orion data.   So, all three data sources  are more powerful when used jointly.

We   define  the  CVSO-VISTA-Gaia   sample  of  PMS   stars  in  section~\ref{sec:sample} and discuss  its properties such as distance, clustering, etc.   Then in  section~\ref{sec:data} the data  on binary stars derived from the combination  of the three surveys are presented and characterized.   The resulting  binary statistics are studied  in section~\ref{sec:stat}. 
We summarize our findings and present our conclusions in section ~\ref{sec:disc}.

%---------------------------------------------------------
\section{The CVSO-VISTA-Gaia sample of PMS stars}
\label{sec:sample}

%-------------------------------------------------
\subsection{Target sample selection}
\label{sec:targ}

The CVSO catalog of young stars in Orion  by \citet{CVSO}  served  as  a starting  point  for our  target selection. The CVSO contains 2062 spectroscopically confirmed T~Tauri stars widely distributed across the Orion OB1 association, mostly in the off-cloud regions, covering well over 100 square  degrees on the  sky \citep[Figure \ref{fig:cvso_vista}; also, Figure  21 of][]{CVSO}.   Though not complete, we consider the CVSO to be a representative sample of the population of PMS K and M type dwarfs in the off-cloud regions of the Orion OB1 association, spanning the OB1a and OB1b sub-associations. First, because of how the sample was selected, it is not biased toward accreting stars with optically thick disks, as would be the case of surveys that select objects with strong H$\alpha$ emission or near-IR excesses. Therefore, we expect it to contain a reasonable representation of both accreting and non-accreting PMS stars, most importantly because the later constitute the bulk of the population in the slightly more evolved off-cloud areas of the association. Second, the spatial distribution of the CVSO PMS stars across the OB1a and 1b sub-associations is uniform enough, and there should be no significant, unexpected biases due to sampling only a small area of one or the other region.  We point out that because of how it was constructed, the CVSO does not represent the much younger, on-cloud population, which we do not address here.
 
Positions reported in the CVSO were determined with a custom pipeline that measured an ($x,y$) weighted centroid for each object; these positions were translated to coordinates on the celestial sphere using astrometric transformation matrices referenced to the USNO A-2.0 catalog \citep{monet1998}.   A positional match between CVSO right ascension and declination coordinates and the 2MASS catalog \citep{2MASS}, using a $1\arcsec$ radius, yields a Root-Mean-Square (RMS) difference of 0\farcs21 $\pm$ 0\farcs17, sufficient for matching each source with other catalogs.

We used TOPCAT \citep{taylor2005} to match the CVSO catalog star positions with the VISTA Orion source catalog,  that provides accurate positions for $\sim 3\,$ million sources referenced to 2MASS (RMS of $\sim 80$ mas\footnote{ http://casu.ast.cam.ac.uk/surveys-projects/vista} for the residual differential astrometry).  The VISTA Orion source positions are an average of the positions determined in each of the photometric bands in which a source was detected. A comparison of the VISTA source positions with the UCAC 4.0 catalog \citep{Zach2013} resulted in  an RMS of $\sim0\farcs27$ for the absolute astrometry, with no systematic offset  \citep{spezzi2015}.  Running a sky match with a $1\arcsec$ search radius between the CVSO and VISTA catalogs yielded 1216 matches, with an RMS of 0\farcs22 $\pm$ 0\farcs17, which is dominated by the errors in the CVSO positions. 

We further restricted the matched sample as follows.  We selected all stars  that spatially belong to  the populations Ori  OB1a or OB1b, as shown in Figure \ref{fig:cvso_vista}, but we excluded a 0\fdg5  radius around the star $\sigma$~Ori,  which is the center of the eponymous stellar cluster. This led to an initial list of 1137 stars, 405 in Ori OB1b, and 732 in Ori OB1a including the 25~Ori and HR~1833 clusters. Table~\ref{tab:sample} contains all targets, numbered sequentially from 1  to 732 and from 1001 to 1407 for the OB1a and  OB1b groups,  respectively. These  internal numbers $N$, along with the  original CVSO numbers, are used throughout the paper. 

%-------------------------------------------------
\subsection{Cross-match with  Gaia and characteristics of the sample}
\label{sec:main}

The next step was to match the sample of 1137 CVSO stars with VISTA catalog  information against the Gaia Data Release 2 catalog (GDR2). We first used Vizier to download all stars in GDR2  within a 30\arcsec ~radius of each of the 1137 CVSO target coordinates. Then, we did a cross-match between this temporary catalog and the CVSO positions, using a 5\arcsec   ~radius and selecting the nearest Gaia source. Coordinate differences between CVSO and Gaia were small for single targets but offsets up to 2\arcsec ~were found for binaries, because CVSO positions refer to their unresolved (blended) images. When the offsets between the CVSO positions and the actual positions of primary components, determined from the VISTA images as  explained below, are accounted for, the rms coordinate difference with Gaia is 0\farcs06. Overall, 1078 CVSO stars have GDR2 astrometry. As for the other 59 objects, 55 of them have no parallax and PM information in GDR2, and 4 had no match at all in the GDR2 for no apparent reason  (these stars are single and of average  brightness:  CVSO 685,  1097, 1283, 1819). Finally, we replaced the CVSO equatorial coordinates with the GDR2 coordinates (equinox J2000, epoch J2015.5), which we use from now on.

\begin{figure}
\epsscale{1.1}
\plotone{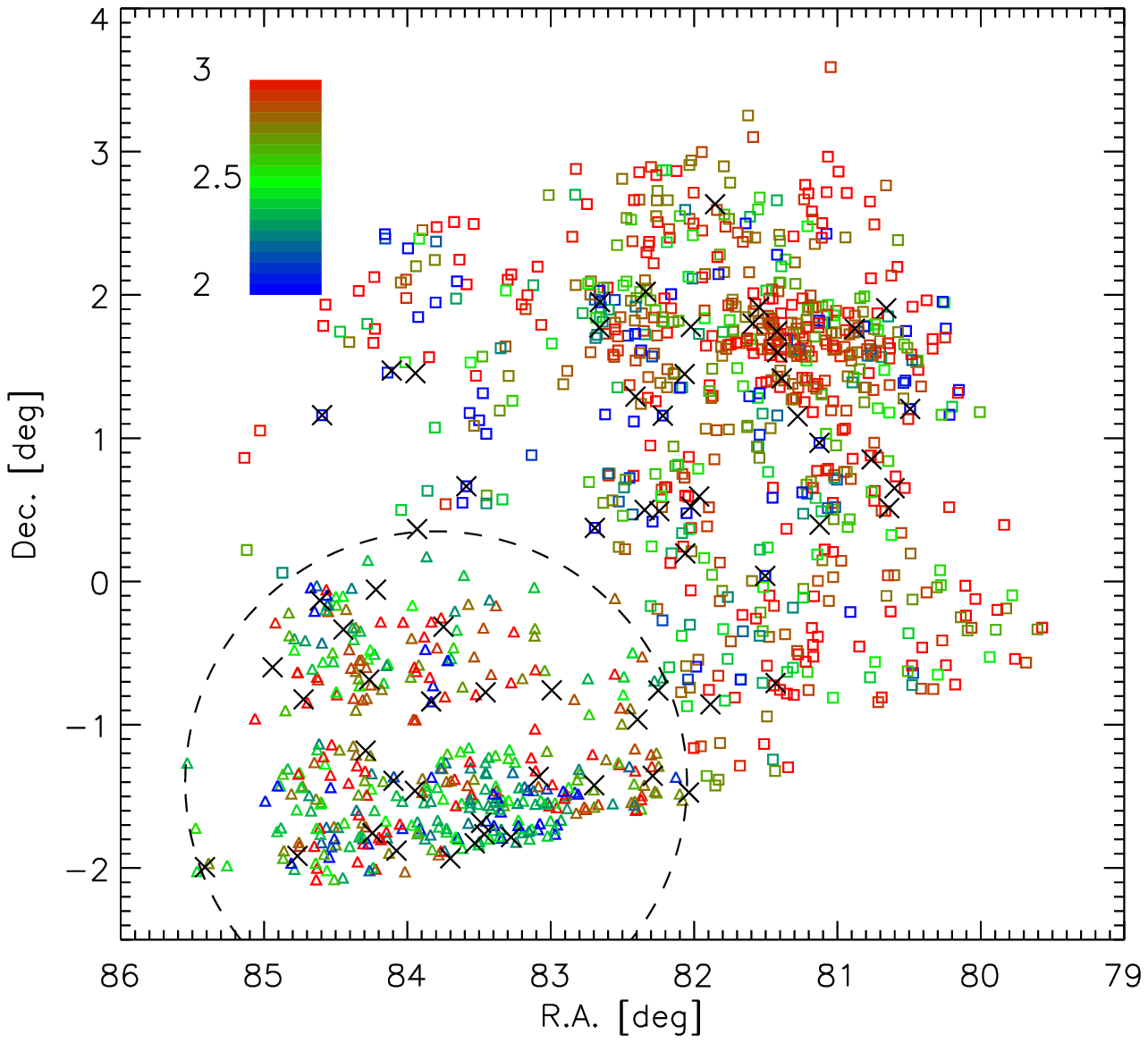}
\plotone{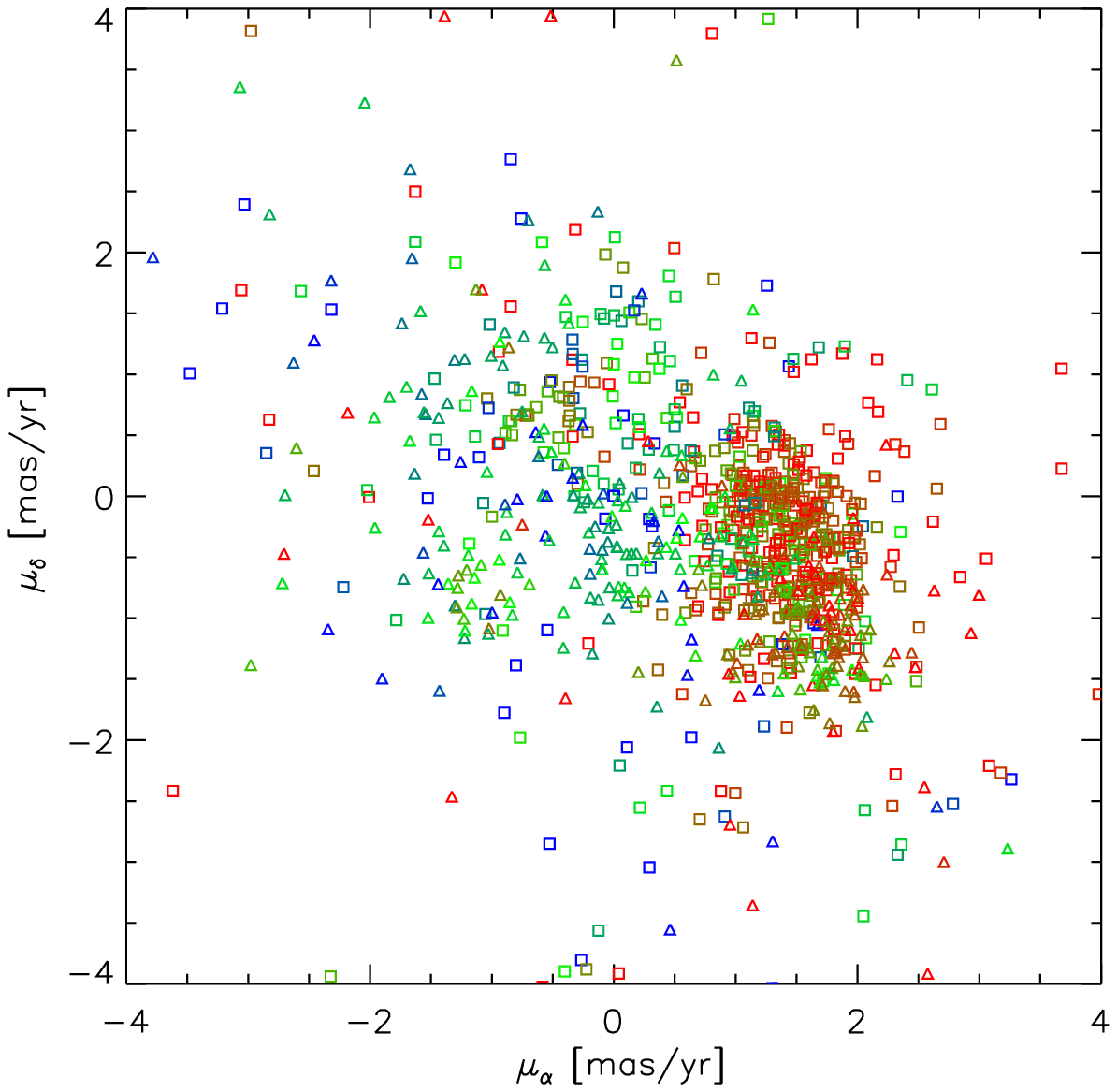}
\caption{Top: location of  the 1137 CVSO-VISTA objects on the sky (squares for OB1a,  triangles for OB1b). The points are colored by parallax in the range  from 2 to 3 mas, as shown  by the color bar. Black crosses are stars  without GDR2 parallaxes. The dashed circle indicates the approximate boundary of the Ori OB1b group. 
Bottom: distribution of the sample  of 1078 objects with GDR2 astrometry in proper motion space. The symbols and colors are  the same as in the top panel.
\label{fig:sky} }
\end{figure}

Having folded in GDR2 astrometry  with the CVSO-VISTA sample, we can now  take a look at the overall astrometric properties of our target sample. The top panel of Figure~\ref{fig:sky} shows the location of the 1137 target stars on the sky,  where the symbols are colored according to the GDR2 parallax. The 55 stars with no parallaxes or PMs in the GDR2 and the 4 missing stars are marked with crosses. The bottom panel of Figure~\ref{fig:sky} shows the PM distribution of  the 1078 targets having  GDR2 astrometry.  The closer stars  (in red) are more  tightly concentrated  in  PM space,  while the  more distant population (green and blue) has a larger PM scatter.  

% plotsample => plotsky ==> plotsky.eps,  plotpm.eps

As already noted  by \citet{CVSO}, the GDR2 astrometry  shows that Ori OB1 stars are mostly located at distances from 300 to 450 pc, depending on  the group.   They also show, in accord with Figure~\ref{fig:sky},  that  the parallaxes  in Ori  OB1b have  a bi-modal   distribution,  indicating   that  closer   stars,  possibly belonging  to  Ori  OB1a,  project   on  the  more  distant  Ori  OB1b group. 

The detailed structure of the Orion star-formation region is complex. It has been the subject of several studies using GDR2 astrometry \citep{Zari2019} and, additionally, radial velocities \citep{Kounkel2018}. Most of our targets belong to the groups C and D identified by Kounkel et al.; these groups have different mean distances (416 and 350 pc, respectively) and radial velocities but spatially overlap on the sky. Since the structure of the Ori OB1 association is outside the scope of this paper, and our focus is on binaries, we use the traditional division into OB1a and OB1b groups based only on the sky location, following the boundaries used by \cite{briceno2005,CVSO}.  Their mean parallaxes are  2.748 and 2.576 mas
 respectively, corresponding to distances of 363 and 388 pc. The mean PMs are close to zero and have a dispersion of $\sim$2 mas~yr$^{-1}$.   We point out that groups OB1a and OB1b as considered here are, however, not homogeneous in terms of their age and distance. OB1a contains clusters like 25 Ori and HR1833 within the more widely spread "field" PMS population. Though it seems clear that OB1a as a whole is a population originating in an earlier star-forming episode compared to OB1b \citep{Kounkel2018,CVSO}, the ages and distances we use are only indicative.

Using GDR2 astrometry, in the next section we will investigate the membership of our targets to Ori OB1a and Ori OB1b, and identify likely non-members. In fact, \citet{CVSO} note that their catalog  of Orion PMS  stars can still be  slightly contaminated by  active foreground K- and M-type  dwarfs with spectral signatures resembling those of  PMS stars. For example, CVSO~569 is a 6\farcs2 pair of similar stars  with almost identical PMs of  $(-20, -7)$ mas~yr$^{-1}$ and  parallaxes about 4\,mas;  this is  a physical  binary, and its spectrum does show H$\alpha$ in emission and Li I 6707 in absorption, therefore it is clearly a low-mass, PMS star but likely foreground and unrelated to the Orion OB1 PMS population.  Finding young stars with motions discrepant from those generally agreed to characterize the bona-fide Orion OB1 population seems increasingly less surprising, since recent studies find that the structure of the stellar population across Orion is richer and more complex than previously thought \citep{chen2019,kos2019}.

%-------------------------------------------------
\subsection{Analysis of the GDR2 astrometry}
\label{sec:DR2}

The large distance to Ori OB1 and its small PM mean that very accurate astrometry is needed to discriminate true association members from foreground and background stars. In addition, unresolved binaries degrade the quality of GDR2 astrometry.  Therefore, we focus in the following on filtering out from our targets the likely non-members, but keep those that potentially have their Gaia astrometry compromised due to the presence of a binary companion. The latter can be evidenced in three different ways.

First, pairs with separations from 0\farcs1 to 0\farcs7 and moderate magnitude difference $\Delta m$ often have undetermined astrometric parameters (parallax and PM)  because they were recognized as non-point sources.  For example, all GDR2 stars without parallaxes were resolved in the speckle interferometric survey of Upper Scorpius \citep{TokBri2019}. There are 55 of our targets that do not have GDR2 parallaxes (section \ref{sec:main}).  Second, the GDR2  astrometry  of many  close binaries,  when present, is  often substantially biased because their  motion does not conform  to the  standard 5-parameter  astrometric  model.  Typically, these  stars have  large errors  of astrometric  parameters,  e.g. the parallax  error  $\sigma_\varpi$.    Our  experience  shows  that  the parameters of  such stars can  deviate from their true  values (known, e.g., from wide components of well-resolved  physical triple systems) much larger than allowed even  by those inflated errors. In short, the GDR2 astrometry of these stars  is unreliable.  Third, even when the 5-parameter astrometric model is adequate,  the PM can still be slightly biased by the orbital motion in a long-period binary. A solar-mass binary with a semimajor axis $a$ (in au) and a typical mass ratio of 0.5 would have the orbital PM on the order of  $1.8(10/a)^{0.5}$ mas~yr$^{-1}$  at a  distance of  370\,pc.  

\begin{figure}
\epsscale{1.1}
\plotone{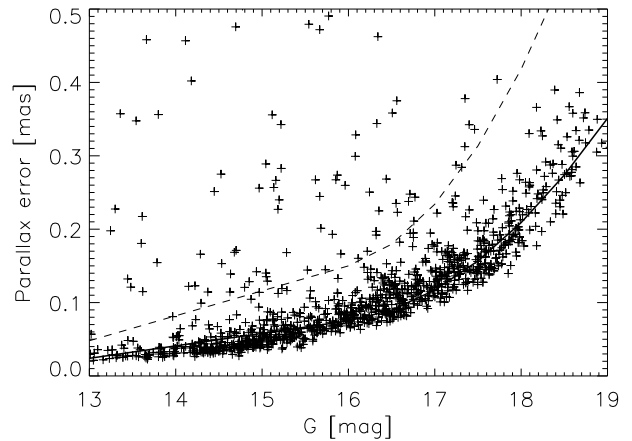}
\caption{Parallax errors vs. $G$ magnitude (crosses). The full line is $\sigma_0(G)$ defined by equation.~\ref{eq:eplx}, the dashed line represents $ 2 \sigma_0(G)$.
\label{fig:eplx} }
\end{figure}
% sample.pro => checkgaia => eplx_model.eps

In order to define a measure for the reliability of GDR2 astrometry specific to our target sample, we plot in Figure~\ref{fig:eplx} the parallax error vs. $G$ magnitude for all targets with GDR2 astrometry  and having $G < 19$. Note, for very faint targets the GDR2 astrometry becomes very uncertain and we therefore reject a priori 17 stars with $G > 19$, which means in the context of our analysis we define those as non-members. The distribution shown in Figure~\ref{fig:eplx} follows a well-defined trend which we approximate by the formula
\begin{equation}
\sigma_0 (G) \approx 0.024 + 0.017 (G-13)  + [0.025(G-16)^2 ]  ,
\label{eq:eplx}
\end{equation}
where the quadratic term is added only for $G > 16$ and $\sigma_0 (G)$ is  in mas.  We  then use the  ratio of  the parallax  error to  its model, $r_\varpi = \sigma_\varpi  / \sigma_0(G)$, as a measure  of the excess astrometric  noise indicative  of  biased GDR2  astrometry.  The  Gaia  errors depend  on  the source  position  on the  sky, and  we caution  against using  our simplistic  model (\ref{eq:eplx})  in a more general context; it is just suitable for Orion.

\begin{figure}
\epsscale{1.1}
\plotone{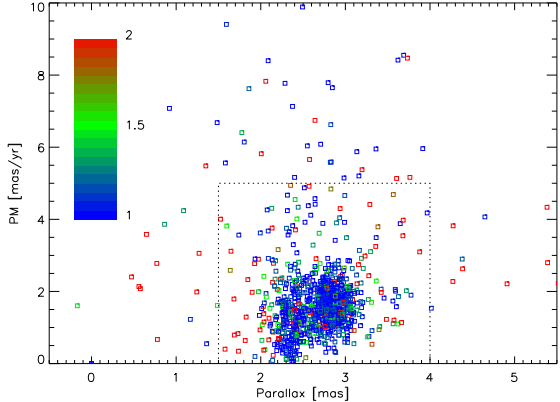}
\caption{Correlation between  parallax and  total PM for our sample.  The  symbols are colored according to the excess  error $r_\varpi$, as shown by the color bar. The dotted box shows the limits of parallax and PM adopted for  the 934  astrometric members. 
\label{fig:pm-plx} }
\end{figure}

Figure~\ref{fig:pm-plx} plots the parallax and total PM of our targets with colors that correspond to $r_\varpi$. Targets with reliable astrometry,  defined here as $r_\varpi < 2$, are tightly concentrated   at   parallaxes  between  2.2   and  3.2  mas  and   PMs  below 3~mas~yr$^{-1}$.   A bi-modal distribution  of parallaxes can be noted. Considering potential biases caused by unresolved binaries, we adopt the following relaxed criteria.  Targets with $G < 19$ and parallaxes from 1.5 to 4 mas and a total PM less than 5\,mas~yr$^{-1}$, irrespective of their excess noise $r_\varpi$, are considered  astrometrically confirmed members of Ori OB1. There are 934 stars that comply with these criteria and are assigned a membership flag 2 in Table~\ref{tab:sample}. The high rate of astrometrically confirmed  members of Ori OB1 validates the spectroscopic and photometric selection of young stars adopted in the construction of the CVSO sample. For comparison, \citet{Kounkel2018} adopted a parallax range from 2 to 5 mas and the PM limit of $\pm$4\,mas~yr$^{-1}$ in both coordinates as membership criteria. 

All targets with reliable astrometry, i.e.\ $r_\varpi < 2$, but total PM and parallax values outside our adopted selection box are considered  astrometric non-members (membership flag 0  in Table~\ref{tab:sample},  89 stars).  The remaining 41 stars with unreliable GDR2 astrometry (likely close binaries) outside the adopted parallax and PM limits (including three with negative parallaxes) are considered as members, unless their total PM is larger than 13\,mas~yr$^{-1}$. This PM threshold is chosen by examining the tail of the PM distribution and applies to only 6 stars, which means 35 stars are finally considered as members. These members are assigned the membership flag 1 to distinguish them from astrometrically confirmed members. Membership flag 1 is also assigned to 52 stars with missing GDR2 astrometry and $G < 19$.   Admittedly, the threshold  $r_\varpi < 2$ used here to define reliable astrometry is  arbitrary; a smaller threshold of 1.5 increases the number of targets with questionable astrometry by 56, but combination of all membership criteria leads to the same sample size of 1021.   

The four stars not found in GDR2, the 17 stars fainter than $G=19$ mag,  and the 6 stars with unreliable GDR2 astrometry and total PM larger than 13\,mas~yr$^{-1}$ are excluded from the following statistical analysis together with the astrometrically confirmed non-members (membership flag 0).  The numbers of targets with various membership status are reported in Table~\ref{tab:members}.    Overall, there are 1021 members, 658 in Ori OB1a and 363 in Ori OB1b. We provide data for all 1137 targets of the original sample and their companions and use the membership flag defined here only for evaluation of  the multiplicity  statistics.

The CVSO-VISTA-Gaia  targets are listed  in Table~\ref{tab:sample}. They are numbered sequentially from 1 to 732 for stars in Ori OB1a and from 1001 to 1407 for those in Ori OB1b (the latter group contains 405 stars).   Within each  group, the  targets are  ordered in  the right ascension.  These  numbers  $N$,  along  with the  CVSO  numbers  from \citet{CVSO}, link the targets to  the lists of double stars presented below.  In the following columns of Table~\ref{tab:sample} we give the information  extracted from  GDR2, namely  the  equatorial coordinates (equinox  J2000,  epoch  2015.5),  parallax $\varpi$,  its error,  proper  motions $\mu^*_\alpha$ and $\mu_\delta$, and the $G$ band magnitude. The $J$ magnitude from 2MASS and the spectral type are retrieved from the CVSO catalog. The  last three columns contain the excess noise  $r_\varpi$ (zero if parallax is  not known), the membership flag, and, for binaries, the separation in arcseconds.

% plotsample => plotsky => pm-plx.eps

%\input{tabsample_small.tex}
% portion of the main table tabsample, see tabsample.pro
\begin{deluxetable*}{rr  cc rrrr rr c ccc}
%\tabletypesize{\scriptsize}     
\tablecaption{CVSO-VISTA-Gaia sample (fragment) 
\label{tab:sample}  }
\tablewidth{0pt}                                   
\tablehead{                                                                     
\colhead{$N$} & 
\colhead{CVSO} & 
\colhead{$\alpha_{2000}$} & 
\colhead{$\delta_{2000}$} & 
\colhead{$\varpi$} & 
\colhead{$\sigma_\varpi$} & 
\colhead{$\mu^*_\alpha$ } &
\colhead{$\mu_\delta$ } &
\colhead{$G$} & 
\colhead{$J$} & 
\colhead{Spectral} & 
\colhead{$r_\varpi$} & 
\colhead{Memb.} & 
\colhead{$\rho$} \\
& & 
\colhead{(deg)} & 
\colhead{(deg)} & 
\colhead{(mas)} & 
\colhead{(mas)} & 
\multicolumn{2}{c}{(mas yr$^{-1}$)} & 
%\colhead{(mas yr$^{-1}$)} & 
\colhead{(mag)} & 
\colhead{(mag)} & 
\colhead{type} &
& & 
\colhead{(\arcsec)}
}
\startdata
   1 & 405 &  79.57240 &  $-$0.32216 & 3.41 & 0.37 &    0.98 &   $-$0.28 &18.56 &14.99 &M4.5  & 1.30 &2 & 0.00\\
   2 & 408 &  79.60478 &  $-$0.33421 & 2.73 & 0.27 &    1.14 &      0.38 &18.56 &15.02 &M3.0  & 0.96 &2 & 0.00\\
   3 & 416 &  79.68400 &  $-$0.56678 & 2.86 & 0.03 &   10.50 &     17.79 &14.73 &12.59 &M0.0  & 0.51 &0 & 0.00\\
   4 & 425 &  79.76144 &  $-$0.54096 & 3.00 & 0.08 &    1.73 &   $-$1.02 &16.07 &13.11 &M3.0  & 1.04 &2 & 0.00\\
   5 & 427 &  79.78025 &  $-$0.09696 & 2.55 & 0.17 &    1.52 &   $-$0.89 &17.62 &14.41 &M3.0  & 1.01 &2 & 0.00\\
   6 & 432 &  79.82293 &  $-$0.18850 & 2.78 & 0.08 &    2.00 &   $-$0.82 &16.51 &13.49 &M4.0  & 0.86 &2 & 3.44
\enddata 
\end{deluxetable*}

\begin{deluxetable}{l c  ccc}
\tablecaption{Classification of the targets
\label{tab:members} }
\tablewidth{0pt}                                   
\tablehead{   
Member flag & $N$   & $0< r_\varpi <2$ & $r_\varpi >2$ & $r_\varpi =0 $ }
\startdata
2  & 934 &  869 & 65 & 0 \\
1 & 87    &  0     & 35  & 52 \\
0 & 116  & 102  &  7  & 7 
\enddata
\end{deluxetable}

%-------------------------------------------------
\subsection{Photometry and CMD}
\label{sec:CMD}

\begin{figure}
\epsscale{1.1}
\plotone{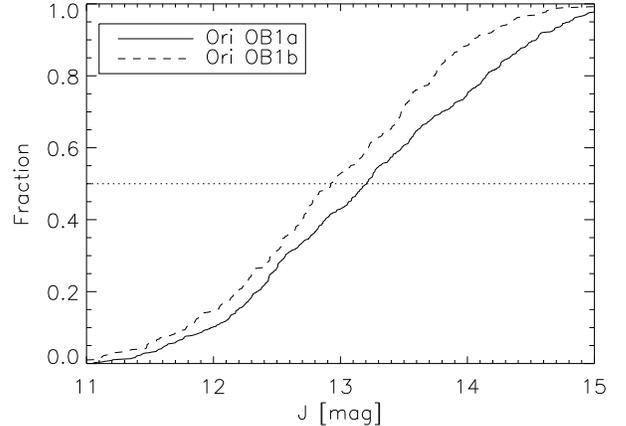}
\caption{Cumulative distributions of $J$ magnitudes  for all targets classified as members (flag 1 or 2).
\label{fig:jmaghist} }
\end{figure}

\begin{figure}
\epsscale{1.1}
\plotone{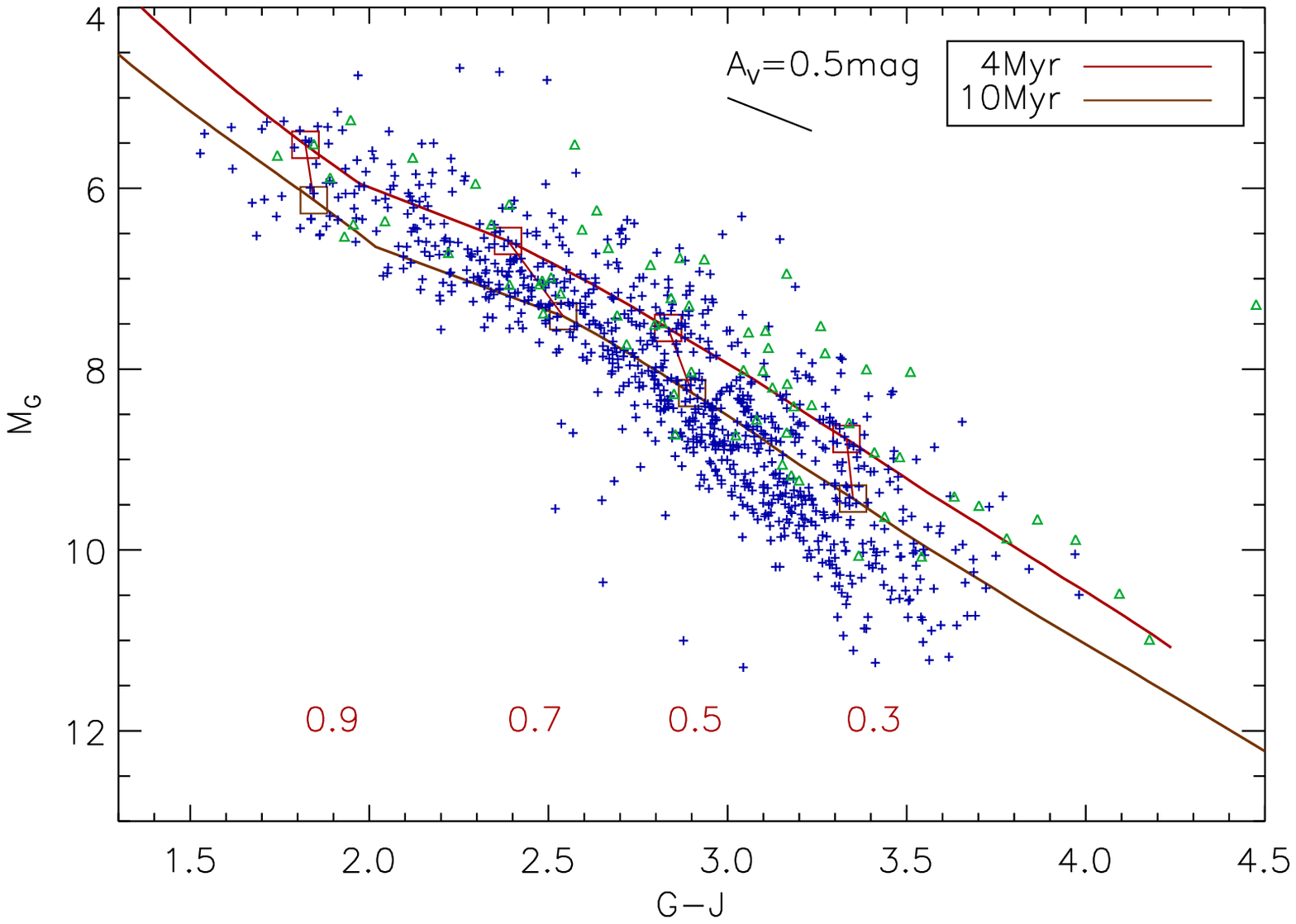}
\caption{Color-magnitude  diagram.   Known  binaries with  separations  less than 5\arcsec  ~are plotted by green triangles,  other stars by  blue crosses.   Absolute magnitudes have been derived  for each star  based on its parallax.   Two PARSEC isochrones for solar metallicity are plotted.  The squares on  the isochrones and numbers mark masses  from 0.3 to 0.9 \msun.  The line marks the effect of an $A_V =0.5$ mag  extinction. 
\label{fig:cmd} }
\end{figure}
% plotsample => cmd

Figure~\ref{fig:jmaghist} shows cumulative  distributions of $J$  magnitudes for members  of Ori OB1a and Ori OB1b. The  medians are 13.21 and 12.93 mag, respectively, consistent with Ori~OB1a being slightly older than Ori~OB1b; the median $G$ magnitudes in these groups are 16.20 and 15.89 mag.

The color-absolute magnitude  diagram (CMD) in  Figure~\ref{fig:cmd} shows only 934 astrometrically confirmed members of the association  with measured parallaxes. We plot  the 4  Myr  and 10  Myr PARSEC  isochrones for  solar metallicity \citep{PARSEC}  using the 2MASS and   Gaia colors and mark corresponding masses;  these ages are consistent with those adopted by \cite{CVSO} for OB1b (5 Myr) and OB1a ($\sim 11$ Myr);  remember though that these groups are not strictly coeval, as noted before.   We do not use  here the VISTA Orion photometry because  for brighter targets it is biased by saturation. The extinction is not corrected for, but since these are off-cloud populations, the overall reddening is small.  In fact, for  our  sample the  median extinction $A_V$  determined in the CVSO catalog  \citep{CVSO} is 0.36 mag. About  24\% targets have $A_V =0$,  and only 11\% have  $A_V > 1$ mag. According to \citet{Danielsky2018},  $A_V = 1$ mag corresponds to $A_G =0.47$  mag and $A_J = 0.24$  mag for a star  of 4000~K effective temperature. The  $A_V = 0.5$ mag vector  plotted in Figure~\ref{fig:jmaghist}  displaces stars almost  parallel to the   isochrones. The low-mass stars appear to be bluer (or fainter) compared to the  isochrones, showing that evolutionary models of PMS stars are still far from perfect.  This systematic deviation from the isochrones is confirmed by our photometry of  binaries, see section~\ref{sec:phys}. 

Binary stars are  located on the CMD above  the single-star isochrone. Known binaries with separations  less than 5\arcsec ~are distinguished in Figure~\ref{fig:cmd}  by green  triangles.  The $G$  magnitudes of those  61 targets  refer to  the primary  components resolved  by  Gaia, while their  $J$ magnitudes from 2MASS refer  to the combined light,  displacing  the points  to  the right  by as much as  0.75 mag. However, the majority of binaries are not recognized because they are closer than 0\farcs6, the resolution limit of our survey. Binarity certainly contributes to the scatter in the CMD.

The  CMDs of  various  sub-groups  of the  Ori  OB1 associations  are plotted and  discussed by \citet{CVSO} and \citet{Kounkel2018}.   They derive model-dependent ages ranging from  4 to 13 Myr for  various sub-groups. However, even within one sub-group the spread of the CMD is substantial. One of the reasons  is that  all CVSO  stars are variable (this  was one  of the selection criteria  in building the sample).  The variability of low-mass PMS stars ranges from a median value of 0.5 mag in the $V$ band for accreting Classical T Tauri stars, caused by a combination of variable accretion, rotational modulation by hot/cold spots and possible disk obscuration, down to $\sim 0.3$ mag for the non-accreting Weak-lined T Tauri stars, in which variability is mainly due to rotational modulation by dark spots and chromospheric activity \citep{CVSO}. The photometry provided in the CVSO  catalog is  averaged  over time,  reducing  the impact of variability.  But the 2MASS and  Gaia photometry are not simultaneous, and they are mostly single-epoch measurements, therefore  variability  contributes  to the  errors  of colors and increases the scatter in the CMD.

Figure~\ref{fig:cmd} implies that most  CVSO stars have masses between 0.3 and 0.9 \msun, with  0.4 to 0.8 \msun ~being dominant, i.e., spectral types $\sim $K2 to M4 \citep[see Figure~1 in ][]{CVSO}.   However, masses  of PMS stars  estimated from absolute  magnitudes or  colors  are known to be  highly uncertain.  The isochrones appear to  deviate systematically from the observed pre-main sequence, and the problem is aggravated  by the intrinsic  variability of all CVSO  stars that  adds uncertainty of magnitudes and colors.    Moreover, ages for individual stars are not well determined and there appears to be a considerable age spread in both sub-associations. Given these intrinsic uncertainties,  we  refrain here from  estimating individual  masses and  mass ratios. A crude estimate of mass ratios based on the isochrones is used here only for the purpose of translating the limit of our survey from photometric contrast into approximate mass ratio.   The isochrones suggest that  the magnitude difference in  the $J$ band  is related to the mass ratio of a young  binary $q = M_2/M_1$ as $q \approx 10^{-0.3  \Delta J}$ (see section~\ref{sec:q}).  In the following, we estimate approximate mass  ratios using this  formula without insisting  on its correctness or  uniqueness. According to this  relation, binaries with $\Delta  J  < 3  (2)$  mag  have $q  >  0.13  (0.25)$.  Therefore,  by restricting our statistical analysis to pairs with $\Delta J < 3$ mag, we cover most  of the mass ratio range,  while rejecting fainter (and mostly unrelated) companions.

%-------------------------------------------------
\subsection{Clustering and chance projections }
\label{sec:cluster}

\begin{figure}
\epsscale{1.1}
\plotone{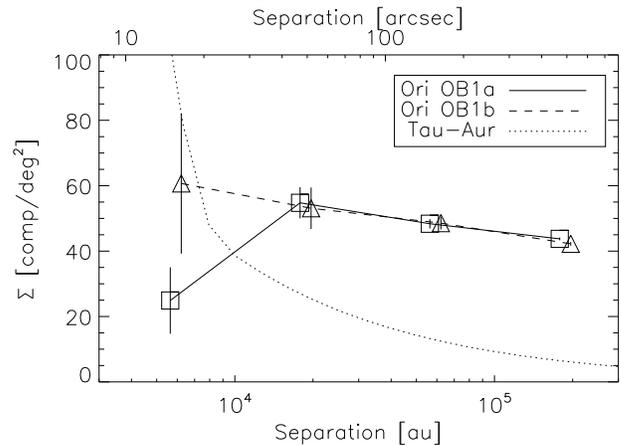}
\caption{Surface density of companions vs. separation in Ori OB1a and OB1b. The dotted line shows the companion density in Taurus  according to \citet{Larson1995}. 
\label{fig:cluster} }
\end{figure}
% clustering.pro => plotcluster 

Companions belonging to the Ori OB1 group according to the astrometric and photometric criteria are not  necessarily bound to the main  targets.  Instead, they could be random pairs of association members projecting close to each other on the  sky.  To elucidate this issue, we computed the  spatial density of  CVSO stars  around each  target in  4 annular zones with  a logarithmic  radius step of  0.5 dex, from  9\arcsec ~to 900\arcsec ~(0\fdg25).   The two subgroups  OB1a and OB1b  are treated separately. The surface  density of  association members  in annular zones  around  our  targets  is plotted  in  Figure~\ref{fig:cluster}, assuming a  common distance  of 370\,pc. The average density  in both groups is similar, about 55 stars per square degree.

The  dotted  line in  Figure~\ref{fig:cluster}  depicts the  companion density  in Taurus  which,  according to  \citet{Larson1995}, is  well approximated by  a broken power law  with the exponents  of $-0.62$ at separations    exceeding   $10^4$   au    and   $-2.15$    at   closer separations. Compared  to Orion OB1,  Taurus has a much  lower density and a  stronger clustering inherited  from the structure  of molecular clouds.  In contrast, in the older Orion OB1 association the stars are well  mixed  at scales  less  than a  parsec,  although  they  retain clustering      at      larger   scales      \citep[][see      also  Figure~\ref{fig:sky}]{CVSO}.

The  first bin  shows a  reduced  stellar density in Ori OB1a   compared to  larger scales,  in strong  contrast with  Taurus. Taken  at face  value, this implies an  anti-correlation, i.e.  avoidance of  close pairs relative to a uniform  distribution.  Most likely,  this is a selection effect   that arises from the construction of the CVSO sample. It used multi-fiber spectroscopy for confirming the PMS nature of $\sim 70$\% of the candidates. Because there is a minimum  distance on the sky between adjacent  fibers \citep[e.g. $20 \arcsec$ for Hectospec;][]{fabricant2005},  close companions (also   PMS   stars)  would not have been   observed   for   this   technical reason. Therefore, we  ignore this effect and assume  that the average density is 55 stars per square degree in both groups.  This means that we  expect to find  1.4 and  5.4 random  pairs of  association members within 10\arcsec  ~and 20\arcsec, respectively, in a  sample of 1021 stars.  However, this is only a lower limit because the CVSO does not contain a complete census of the association members; this is further explored below using GDR2. The expected  number of  random  pairs is  subtracted in  the following analysis  of the separation  distribution.    We restrict the statistical analysis to separations below 20\arcsec ~and to moderate $\Delta m$ to minimize the impact of  random pairs.  Extending  these limits would  aggravate the uncertainty caused by random pairs of association members.

%---------------------------------------------------------
\section{Observational data and their analysis}
\label{sec:data}

Our  primary source  of data  on binaries  is the  examination  of the images from the VISTA Orion mini-survey.   We  attempted to detect almost all  companions within 7\arcsec  ~from  all original 1137 CVSO stars  visible in   the  images and  only later  realized  that the  detection depth  is   excessive for our survey that needs  only a contrast up to 3 mag.  When studying the binary frequencies (section~\ref{sec:stat}) we will restrict the analysis to the   members of Ori OB1. We  complemented the image analysis by searching for  wider  pairs  in  the  VISTA Orion  photometric  catalog  and  by identifying all pairs in the GDR2.  Joint analysis of this information allows  us  to discriminate  real  binaries  from unrelated  (optical) asterisms and sets the stage for the statistical analysis presented in section~\ref{sec:stat}.

%-------------------------------------------------
\subsection{Detecting binaries in the VISTA images}
\label{sec:images}

\begin{figure}
\epsscale{1.1}
\plotone{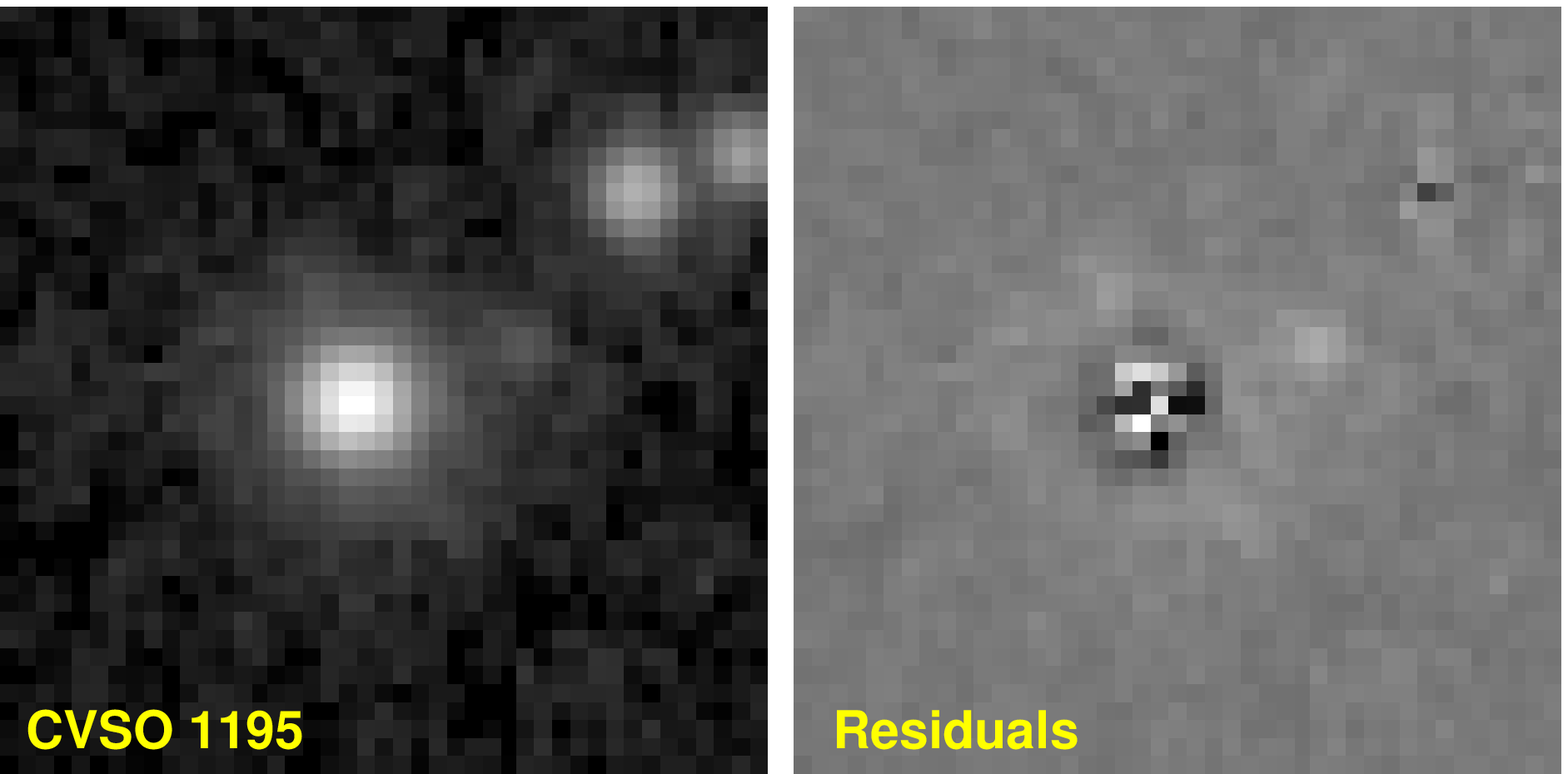}
\plotone{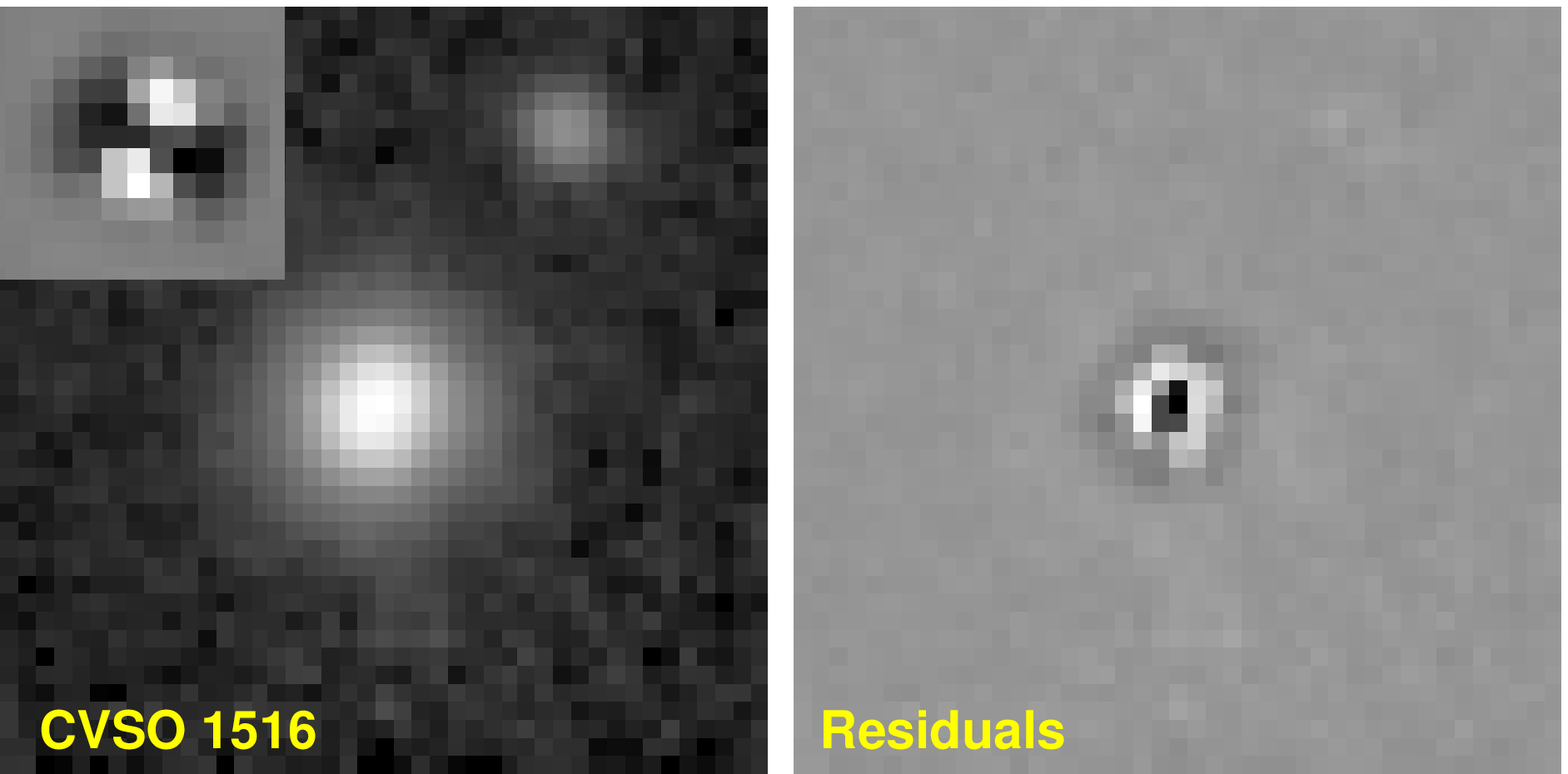}
\caption{Modeling the  postage-stamp images by  Moffat functions. The  images are shown on the left,  the residuals on the right.  Top: the  image of target 1027  (CVSO~1195) in the $H$ band  with three stars. Note another  faint star in-between  that has  been ignored.  Bottom: the  image of target 1176 (CVSO 1516) in  the $J$ band, with an insert showing  residuals for the main  star indicating the elongation. The residuals after fitting three stars (on the right) are smaller.
\label{fig:fit} }
\end{figure}
% CBSO1195.fig (GR.1027)

The VISTA Orion mini-Survey \citep{Petr2011} provides seeing-limited images with a typical  FWHM resolution of 0\farcs9 (details are given at the end of this section). Data  were acquired with VIRCAM,  the VISTA nIR wide-field imaging camera that has an average pixel scale of 0\farcs341/pix. Observations at $Z,Y, J, H, K_s$ bands for one field  were executed sequentially in all filters, spanning no  more than  2--3 hours   in total.  This way effects of variability on the stars' colors should have  been diminished.  

For each CVSO target, fragments of VISTA images of 43$\times$43 pixels, corresponding to a size of 14\farcs7$\times$14\farcs7,  centered  on the  nominal  target position  reported in  the CVSO catalog  were  selected.   They are  called ``postage  stamps''  or  ``stamps''.  Companions  with  separations   up  to 7\arcsec  ~(up  to 10\arcsec  ~in the  corners)  can  be found  in  these postage  stamps. The VISTA/VIRCAM focal plane is  a mosaic of detectors, with large gaps in between, that must be dithered in a 6-point pattern to contiguously fill the field  of view. Furthermore, at $Z$ and $Y$ filters long and short exposures were taken. This means targets were  imaged  several times  in all  five filters.  However,  images where targets  fall near a  detector edge are partially truncated.  Among the 37,464 postage stamps used in this project, 836  severely truncated ones  are ignored. On  average, there are 5 images per  target in the filters $Z$ and $Y$,  10 images in $J$ and $H$,  and only 2.4 in  $K_s$; some targets lack  the $K_s$--band images altogether.

We used  a custom  IDL code to  process these images,  automating the work as  much as possible.  For  each target, the  program selects all postage stamps and displays the  chosen (usually the first) image in a graphical window.   The user defines  the number of visible  stars and their  approximate  positions and  fits  a  model  to  determine accurate relative  positions and intensities  of these  stars.  Modeling  of all other images  of this  same target  can then be  done by  one command, using previous results as a first approximation.

Several  important comments  are in order  here. First,  all CVSO targets are  4 to 6 mag brighter than the faint magnitude limit of  the VSTA Orion survey, assuring that well-resolved companions with a contrast under  3    mag   are    always securely  above    the    noise-limited   detection  threshold. Second,  no good estimates  of the point  spread function  (PSF)  are available because  many postage  stamps contain  just one  star, the target itself. So, a decision on whether the PSF asymmetry is  caused by  a  close  semi-resolved companion  or  by a  residual  telescope  jitter (the  ellipticity of  most images  reported in  the headers  is under 0.05,  but in some  cases reaches 0.1)  is not always  straightforward.  Unlike  the situation  in the survey of \citet{defurio2019}, where accurate models of both PSF and noise were  available, detections of  close companions in  the VISTA  images  cannot  be  automated   and  their  significance  cannot  be  rigorously  evaluated by a  metric like  $\chi^2$.  On  the positive  side, however, we have multiple images of each target obtained under  different  seeing  conditions   in  five  filters.   Therefore,  the  companions are confirmed  as many times as there  are images. Mutual  agreement  of binary-star parameters  derived from  many independent postage-stamp  images guarantees the reliability  of detections; they  are all secure  and there are no  false positives,  as indicated by the independent detection of all, except one, VISTA close binaries (0\farcs6 $< \rho  <$ 1\farcs2) by GAIA and/or high spatial resolution observations (cf.\ section~\ref{sec:SOAR}). The detection limit is further discussed in section~\ref{sec:det}.

The background  level in each image  is determined by  the median pixel value  and further  refined by  excluding pixels  around  known stars within a radius of four times the FWHM resolution, typically about 12 pixels.  Images of  stars that do not  overlap significantly can  be modeled by fitting a symmetric Moffat profile
\begin{equation}
F(x,y) = \frac{p_2}{[1 + r^2/a^2]^{\beta}}, \;\; r^2 = (x - p_0)^2 +
  (y - p_1)^2 
\label{eq:Mof}
\end{equation}
with five free parameters  $p_0,p_1,p_2,p_3=a,p_4 = \beta$.  The first two parameters are  pixel coordinates of the center,  the third is the maximum intensity, the parameters $a$ and $\beta$ define the width and shape  of the  point spread function  (PSF).  The  FWHM equals  $2 a \sqrt{ 2^{1/\beta} -  1}$.  The background level  is  subtracted  prior  to  fitting  and  not  included  in  the model. The PSF is fitted by minimizing $R$, the un-weighted normalized rms difference  between the image $I_i$  and its model  $M_i$ over all pixels $i$ within a radius of 10 from the center:
\begin{equation}
R = \sqrt{ \sum_i ( I_i - M_i)^2} / \sqrt{ \sum_i I_i^2} .
\label{eq:R}
\end{equation}

The  residuals for single stars  are dominated by  the difference between the actual PSF  shape and its model (\ref{eq:Mof}),  rather than by  the detector and photon noise.  Hence $R$ is the appropriate goodness  of fit  metric and its minimization achieves  the best approximation  of the  PSF shape.  For modeling saturated stars, pixels near the center can be excluded. We  also   tested  elliptical   Moffat  models  with   two  additional parameters, ellipticity  and orientation, but found  that the  symmetric model (\ref{eq:Mof}) works well in most cases;  therefore, the elliptical Moffat function was not used. 

When the pair is well separated, we model the secondary star by fixing the  PSF parameters  to  those of  the  primary and  fit only  the position and relative intensity.   For partially overlapping stars, we have the option of adjusting the common parameters $a$ and $\beta$ for all stars, i.e.  $2+3n$ parameters  for an image containing $n$ stars. This method works very well  even for close (blended) pairs  and  it was used for measuring all companions.   Residuals after  fitting  a  triple  source  1027  (CVSO~1195)  are  shown  in Figure~\ref{fig:fit}. The  two companions are separated  from the main star by 6\farcs7 and 8\farcs7 and  have $\Delta J$ of 2.3 and 2.9 mag, respectively; both are unrelated field stars.

Detection  of close  binaries with  separation less  than the  FWHM is helped  by  modeling  the  PSF  by a  symmetric  Moffat  function  and visual examination of the residuals.  A persistent asymmetry of multiple images of the  same target indicates a real  companion, as opposed to occasional PSF elongation. An {\it a posteriori} test of companion detection    is   furnished    by   comparison    with   Gaia  (section~\ref{sec:Gaiabin}). The  lower panel  of  Figure~\ref{fig:fit}  illustrates  modeling of the close binary star 1176 (CVSO 1516)  in an image with a FWHM resolution of 0\farcs96.  The residuals after approximating the central star by a symmetric Moffat  function (in the insert) have  a ``butterfly'' shape indicative of  asymmetry and are large,  $R= 0.156$. Modeling  the central star by two point sources separated by 0\farcs61 yields smaller residuals of $R=0.056$.  This  pair  and the 0\farcs75 pair \# 697 (CVSO 1567) were  overlooked initially  in the VISTA  images  but found  in    Gaia  and re-fitted.   All  other overlooked Gaia pairs are closer than 0\farcs6.

%\input{tabbin_small.tex}
% portion of the main table tabsample, see tabsample.pro
\begin{deluxetable*}{rr  rrr  ccccc  ccccc}
\tabletypesize{\scriptsize}     
\tablecaption{Companions found in the images (fragment) 
\label{tab:bin}  }
\tablewidth{0pt}                                   
\tablehead{                                                                     
\colhead{$N$} & 
\colhead{CVSO} & 
\colhead{$\theta$} & 
\colhead{$\rho$} & 
\colhead{$\sigma_\rho$} & 
\colhead{$\Delta Z$} & 
\colhead{$\Delta Y$} & 
\colhead{$\Delta J$} & 
\colhead{$\Delta H$} & 
\colhead{$\Delta K_s$} & 
\colhead{$\sigma_{\Delta Z}$} & 
\colhead{$\sigma_{\Delta Y}$} & 
\colhead{$\sigma_{\Delta J}$} & 
\colhead{$\sigma_{\Delta H}$} & 
\colhead{$\sigma_{\Delta Ks}$} \\
& & 
\colhead{(degr)} &
\colhead{(\arcsec)} &
\colhead{(\arcsec)} &
\colhead{(mag)} &
\colhead{(mag)} &
\colhead{(mag)} &
\colhead{(mag)} &
\colhead{(mag)} &
\colhead{(mag)} &
\colhead{(mag)} &
\colhead{(mag)} &
\colhead{(mag)} &
\colhead{(mag)} 
}
\startdata
   1 & 405 &340.1 &3.653 &0.201 &  5.61 &  5.59 &  5.06 &  4.89 &\ldots &  0.19 &  0.35 &  0.16 &  0.00 &\ldots\\
   4 & 425 &154.6 &3.457 &0.019 &  5.06 &  5.12 &  5.24 &  5.23 &  4.93 &  0.15 &  0.24 &  0.00 &  0.00 &  0.00\\
   5 & 427 & 12.1 &7.249 &0.026 &  2.64 &  3.23 &  3.57 &  3.88 &  3.76 &  0.02 &  0.16 &  0.16 &  0.30 &  0.08\\
   6 & 432 &101.9 &3.440 &0.019 &  2.15 &  1.91 &  1.76 &  1.89 &  1.76 &  0.03 &  0.04 &  0.00 &  0.04 &  0.01\\
  14 & 455 &228.5 &7.596 &0.147 &  5.58 &  5.59 &  5.73 &  5.70 &\ldots &  0.19 &  0.31 &  0.00 &  0.00 &\ldots\\
  19 & 461 & 40.5 &7.402 &0.041 &  4.62 &  4.64 &  4.14 &  3.94 &  3.16 &  0.13 &  0.24 &  0.15 &  0.07 &  0.07 
\enddata 
\end{deluxetable*}

Although our statistical analysis considers only companions with a contrast up to 3 mag, for the sake of completeness  we report in Table~\ref{tab:bin}   all companions with $\Delta m  < 7$ mag  found  in the postage stamp  images.  Its first  two columns  give  the  sequential  and  CVSO numbers  matching  those  in Table~\ref{tab:sample}.    The position  angle $\theta$  and separation $\rho$  are average  values for  all processed  images in  all filters where the  given companion is detected. The  rms scatter $\sigma_\rho$ gives   an   idea   of    the   internal   agreement   between   these measurements. The  following five  columns give the  average magnitude differences in the  VISTA $Z, Y, J, H, K_s$  bands. The remaining five columns  contain the  rms scatter of $\Delta m$  in each  filter where  two  or more measurements are  available. For a single  measurement, the scatter is  zero. Some  companions lack  measurements in  some  filters either because these  images are unavailable  or because the  companions were not  detected  owing  to  noise  or  truncation.   Table~\ref{tab:bin} contains  490 rows, i.e.\ unique companions to our 1137 targets, with all companions having a detection in at least two filters. The  majority of targets have one companion, and at most four. Most of these companions are unrelated field stars. 

The Moffat models also  provide the FWHM resolution in each image  through parameters $a$ and  $\beta$.  Its median value is 0\farcs88,  the mean  is 0\farcs89, and  the dispersion is 0\farcs20.  Ninety per  cent of FWHM values are comprised between 0\farcs74 and 1\farcs12.

%-------------------------------------------------
\subsection{Wide pairs in the VISTA  Orion photometric catalog}
\label{sec:ptm}

The VISTA Orion  photometric catalog contains  equatorial coordinates and $Z,Y,J,H,K_s$ magnitudes of all  point sources found in the survey.  The catalog is typically complete  (at $10\sigma$ significance) to $Z=21.7$, $J=19.6$, and $K_s=17.9$ mag, as inferred from the histograms of  the magnitudes. This ensures that the companion search in the catalog is sensitive to all companions with $\Delta  J < 3$ mag, as the faintest target has $J\sim16$ mag. 

The catalog was also used to study the stellar density as a function of magnitude in the area of the Ori  OB1 association, to account statistically  for the background contamination. The number of stars $n$ per  square degree brighter than a  certain $J$ magnitude is $\log_{10}n(J) \approx 3.44 + 0.24(J -15)$ in both groups of Ori OB1.  These models are  no longer needed in the light of  Gaia, but could be used to compute the density of unrelated companions. 

We retrieved as companions all catalog stars that are separated between $2\arcsec-20\arcsec$  from the CVSO target position and have $\Delta J < 3$ mag. Separations and position angles of these wide pairs are deduced from the equatorial coordinates. We compared the results of our postage-stamp processing with the VISTA Orion  catalog for  pairs wider  than $\sim$2\arcsec.   The comparison revealed that  coordinates of single  stars in the CVSO  catalog match their VISTA Orion  coordinates with a  median offset of only  0\farcs047 and the  maximum offset of  0\farcs23. However,  for binaries  the spatial match was much worse   because the  CVSO positions  refer  to the  centroids  of  blended  images owing to  its  3\arcsec  ~typical FWHM  resolution.  Our image analysis  gives the offsets   of the main component from the postage-stamp center (which is at the  CVSO  position). The  CVSO coordinates  corrected for  these offsets  match  the VISTA positions  with a  median difference  of 0\farcs05.  The improved target coordinates    also   help  to  match   the  main  stars  to the  GDR2  without ambiguity. Overall,  we found a  very good agreement  between the relative astrometry  and  photometry of  common  pairs  produced  by our  image modeling with those derived from the VISTA Orion catalog, although there are a few  outliers. Companions wider than  7\arcsec ~are  found in the VISTA Orion catalog, without the need to examine the images.  We added  411 wide pairs with $\Delta J < 3$ mag and separation 7\arcsec $< \rho  < 20\arcsec$ to the companion list, thus extending the separation range to 20\arcsec.  The  majority of  these wide companions  are unrelated field stars, as  shown below in section~\ref{sec:phys}. 

%-------------------------------------------------
\subsection{Companions in GDR2}
%\subsection{Double stars in GDR2}
\label{sec:Gaiabin}

\begin{figure}
\epsscale{1.1}
\plotone{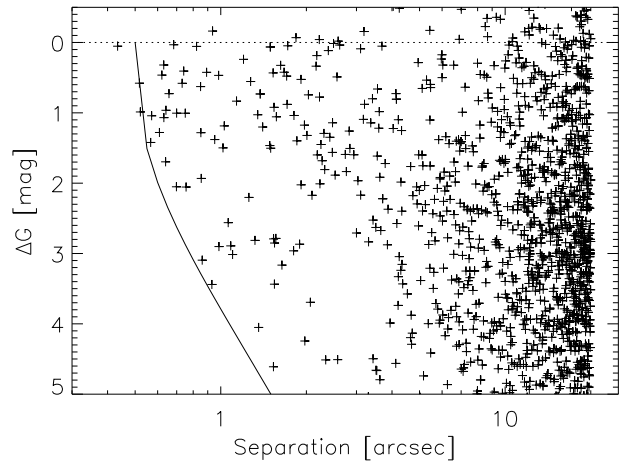}
\caption{Separation versus  $\Delta G$ of companions found  in  Gaia  within 20\arcsec.   The line  shows the   detection limit $\Delta G < 5(\rho - 0.5)^{0.4}$   that approximates the  detection limit  at 50\% probability found by  \citet{Brandeker2019}.
\label{fig:gaiacomp} }
\end{figure}
% gaiacomp.pro => plotgaiacomp

The GDR2 catalog was queried within 30\arcsec ~radius of each target. The coordinate offsets determined from the image analysis helped to match securely the main targets with GDR2 (the  coordinates agree within $\sim$0\farcs06). 

Most  stars found in GDR2  around each target are unrelated (optical) companions. Their separations  and $\Delta G$ are plotted in Figure~\ref{fig:gaiacomp}.  We matched  the VISTA  Orion pairs  to the list of companions in GDR2  and thus retrieved the GDR2 astrometry and photometry for all pairs, except for four close ones with separations below $\sim$0\farcs7  (targets No.\ 392,  597, 1071, 1270)  not resolved  by  Gaia (possibly their secondary components were too faint in the $G$ band).  Conversely,  six close pairs in GDR2 were not recognized in the VISTA images (targets No.\ 253, 458, 697, 1176 1257, 1258).  All   except  two  have  separations  below  0\farcs6.   Sources  697   (0\farcs75)  and  1176  (0\farcs61)  were  overlooked in  the original analysis  of the VISTA images  and added  later (see  Figure~\ref{fig:fit}).  All other companions found in GDR2  with separations $>$0\farcs6 were also detected in the VISTA images or in the survey catalog. 

The  membership  in  the  Ori  OB1 association  was  tested  for  each companion candidate with  reliable  astrometry   (measured  parallax  and  PM  and $r_\varpi  <  2$).   If its  parallax  and PM  satisfy the membership criteria adopted here (Figure~\ref{fig:pm-plx}),  the pair is considered real  (physical)  and  its  flag  is  set  to  $p_{\rm  astro}  =  1$. Otherwise, $p_{\rm  astro} =  0$ and the  pair is  considered optical. For   78  companions with  missing  or  unreliable  astrometry, we  set $p_{\rm astro} = 0.5$ and  use other criteria to test their membership (see section~\ref{sec:phys}).  The rms PM  difference between the primary and secondary components  of 40  physical  pairs  with $p_{\rm  astro}  = 1$,  $\rho <5\arcsec$,  and  reliable  GDR2  data  for  both  components  is  0.6 mas~yr$^{-1}$.     We  suspect that  unrecognized inner  subsystems contribute  to the  scatter of  relative  motions in these wide pairs.

\begin{figure}
\epsscale{1.1}
\plotone{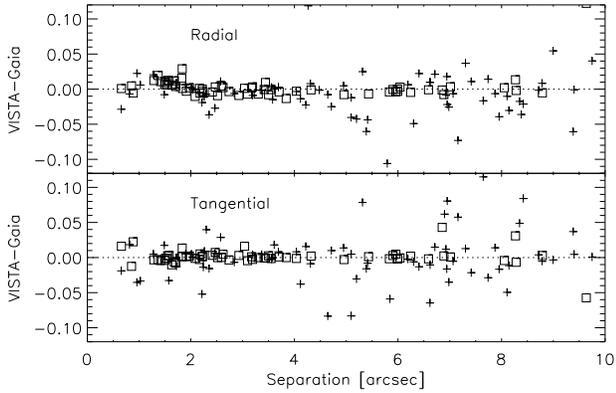}
\caption{Comparison of double-star  astrometry between VISTA Orion and  Gaia.   Squares  are  astrometrically  confirmed  association  members, pluses are other (mostly optical) companions with $\Delta G  < 3$  mag. The top plot  compares separations, the  lower plot shows  the tangential difference $\rho  \sin \Delta \theta$. The axis scale  is in arcseconds.
\label{fig:compastro} }
\end{figure}
% comp.pro => compastro

We compared the relative astrometry derived from the VISTA images with the presumably more accurate GDR2 astrometry (Figure~\ref{fig:compastro}). The agreement is excellent  for astrometrically confirmed members with $p_{\rm astro}  = 1$ and $\rho <  7''$, measured by us  in the images. The  mean offsets between  the relative  companion's positions  in the VISTA images and in  Gaia in the radial and tangential directions are  +1 and  +2  mas, respectively,  while  the rms  scatter of  these offsets  is  8\,mas  in  both  directions.  On  the  other  hand,  the positions of pairs  wider than 7\arcsec ~rely on  the coordinates from the VISTA  Orion catalog  and are  accurate only to  a fraction  of an arcsecond.

%-------------------------------------------------
\subsection{Detection limit}
\label{sec:det}

\begin{figure}
\epsscale{1.1}
\plotone{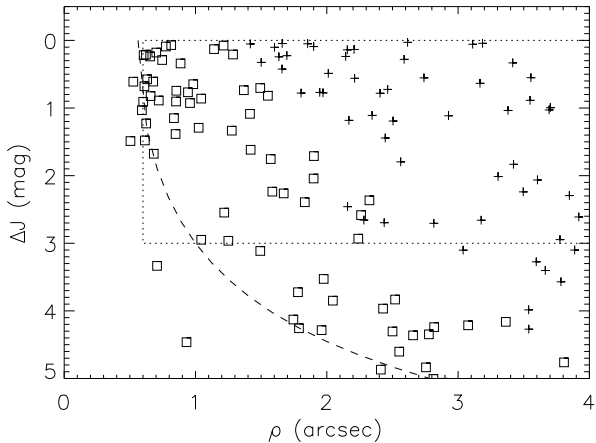}
%\plotone{fig8.eps}
\caption{Location   of   companions   in   the  $(\rho,   \Delta   J)$  plane.  Squares  denote pairs  found in our image analysis, crosses  plot  companions found  both in the images  and in the  VISTA Orion catalog. The  dashed  line is the  empirical detection  limit (eq.\ \ref{eq:detvista}),  the dotted rectangle marks the limits of our statistical analysis.
\label{fig:det} }
\end{figure}
% vistacomp.pro => vistastat

Figure~\ref{fig:det}  plots  the location  of  all  companions in  the $(\rho, \Delta  J)$ plane.   One can appreciate  the advantage  of our image analysis in terms of  resolution and contrast, compared to using solely the VISTA Orion catalog.

The companion  detection in the  VISTA images depends on  the variable FWHM resolution  and on the  signal to noise  ratio.  Here we  use the simplified  optimistic  empirical  detection  limit  (dashed  line  in Figure~\ref{fig:det}) described by the formula 
\begin{equation}
\Delta J < 6 (\log \rho + 0.25)^{0.5},  \;\;\; \rho > 0\farcs6 .
\label{eq:detvista}
\end{equation}
This  formula  is  chosen  ``by  eye''  to fit  the  envelope  of  the points.   We also studied  the empirical detection threshold  as a function of  FWHM resolution, but, considering the  limited range of  FWHM  variation  and  multiple  images available  for  each  target,   decided on  a simpler alternative  (\ref{eq:detvista}). According to  this  formula,  all pairs  with  separation  $\rho  > 1\farcs2$  and  $\Delta J < 3$ mag are  detected in the images. In the following, we  restrict the statistical  analysis to pairs with $\Delta  J < 3$ mag  and  ignore   fainter  companions,  making   their  detection  limit  irrelevant to our study. Only  the contrast and resolution limit at small separations is relevant.  

\begin{figure}
\epsscale{1.1}
\plotone{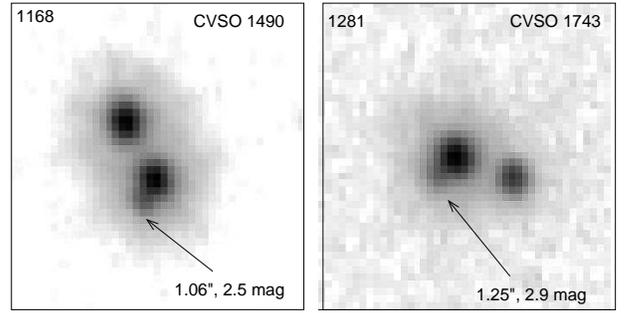}
\caption{   Postage-stamp  $H$-band images  of two  triple systems    with close  faint companions, in  negative rendering.  Separations   and $\Delta  J$ of close  inner pairs are indicated.   Left: target    1168 (CVSO  1490), FWHM resolution 0\farcs73.   Right: target 1281    (CVSO 1743), FWHM resolution 0\farcs87. 
\label{fig:triples} }
\end{figure}

As noted above, all our detections  with $\rho > 0\farcs6$ are secure (no false positives). However, visual detection of close companions by examination of residuals after subtracting the Moffat profile is subjective. Comparison with Gaia reveals that two close pairs  were actually missed by our subjective procedure; they were recovered later by modeling images as a double, rather than single, source. Conversely, four similarly close pairs were unresolved by Gaia.  The mean $\Delta J$ of physical companions in the 0\farcs6-1\farcs2 and 1\farcs2-2\farcs4 separation bins are 0.86 and 0.95 mag, respectively. 

To further probe  the adopted detection limit, we  examined five pairs with $\rho <  0\farcs7$ and $\Delta J > 1$ mag  (targets No. 392, 565, 674,   689,  and   1366),  near   or   beyond  the   dashed  line   in Figure~\ref{fig:det}.  Only the first and most difficult one (No. 392, 0\farcs50, 1.4  mag) was  undetected by Gaia.   This pair  is measured from 2 to  4 times in each VISTA band (12  measurements in total) with consistent parameters (rms separation scatter of 0\farcs08), hence its detection is  highly significant.  We also re-examined  four close and high-contrast pairs  with $1''  < \rho <  1\farcs5$ and $\Delta  J > 2.5$ mag, near  the lower-left corner of the  relevant parameter space (targets No.  1042, 1168,  1190, and 1281).  All these  companions are resolved by Gaia.  Two targets illustrated in Figure~\ref{fig:triples} are triple  systems where the   1\arcsec   ~high-contrast pairs are  accompanied by  wider and  brighter companions  at $\sim$3\arcsec ~separation,  also members of  the association  according to  the GDR2 astrometry. These young triple systems with comparable separations may be interesting in  their own right. They are shown  here to prove that their close  high-contrast inner pairs  are quite obvious and  hard to miss.

An external test of our detections  is furnished by Gaia. We selected all GDR2 companions to our targets  with $0\farcs6 < \rho < 20\arcsec$ that conform  to our  astrometric criteria  of  membership in  Ori OB1  and cross-matched  them  with our  list  of  companions  derived from  the VISTA Orion survey. No missed  pairs closer than 7\arcsec ~were found, apart from  the two close ones  mentioned above. We  conclude that the number of  missed pairs (false negatives)  is very small  or zero. The joint use  of two independent surveys, VISTA-Orion  and GDR2, produces high-confidence results. 

%-------------------------------------------------
\subsection{Discrimination between physical and optical pairs}
\label{sec:phys}

The GDR2 astrometry of  78 companions with $p_{\rm astro} = 0.5$ is either not available or unreliable.  Their membership status is decided based on  the photometry  and  other  criteria and  formalized  by the  flag $p_{\rm phys}$.  For the  majority of other companions with good astrometry, $p_{\rm phys} = p_{\rm astro}$ takes the values of either 1 or 0.

\begin{figure}
\epsscale{1.1}
\plotone{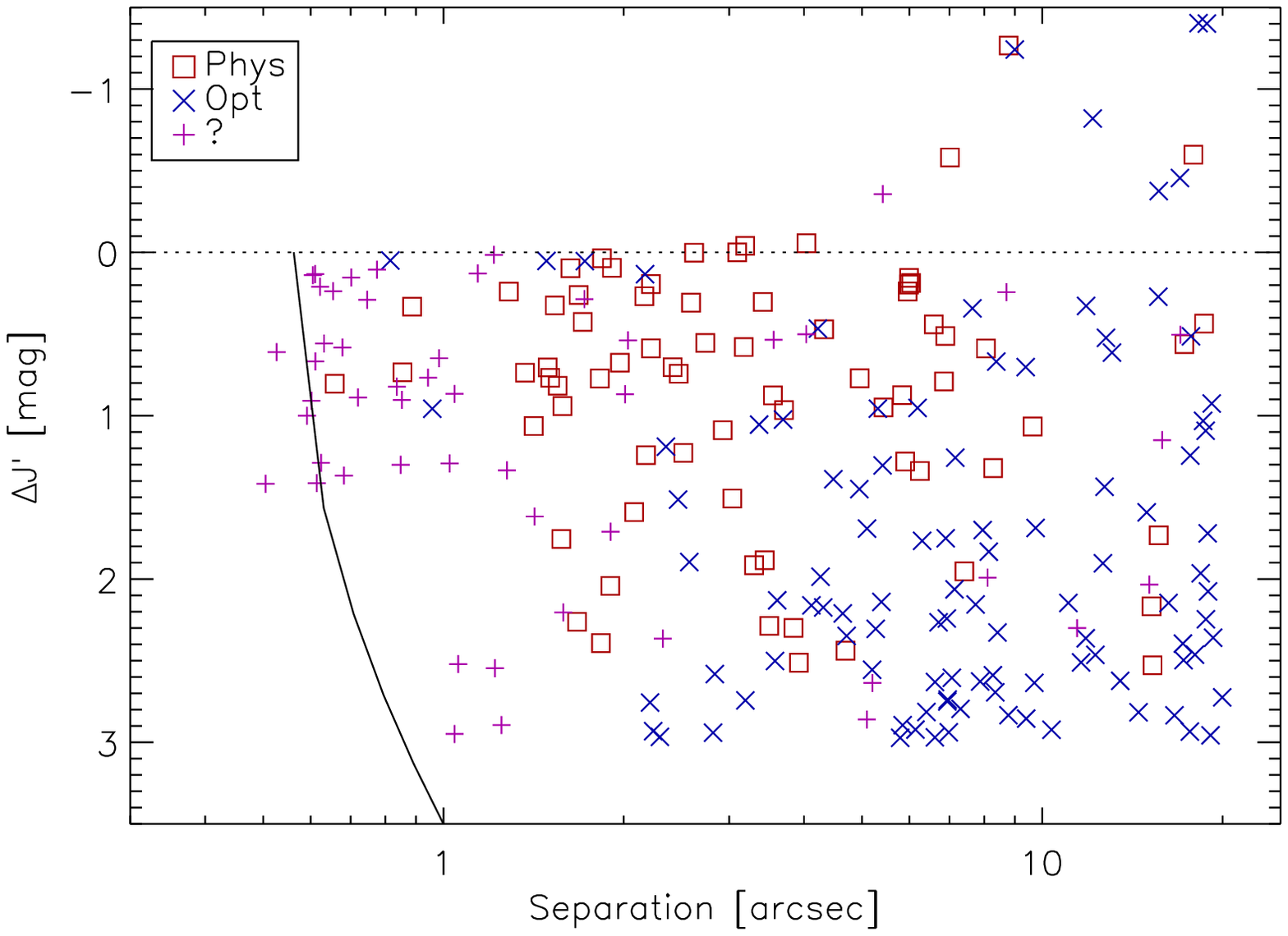}
\plotone{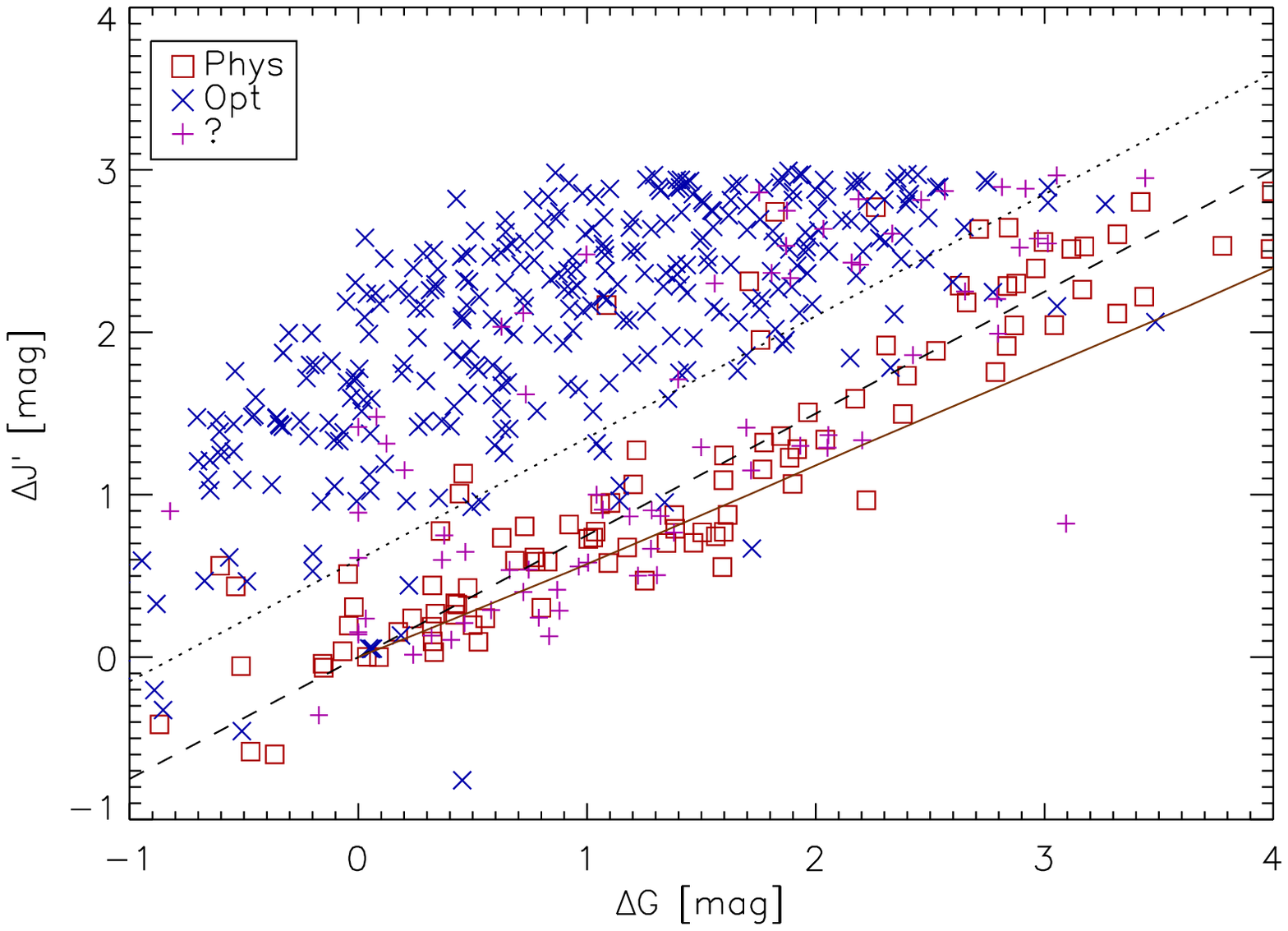}
\caption{Distribution of  companions in  the $(\rho, \Delta  J')$ plane  (top)    and    in    the    $(\Delta   G,    \Delta    J')$    plane   (bottom).  Astrometrically confirmed companions  are plotted  by red   squares, non-members by blue  crosses, and uncertain companions with  $p_{\rm astro} = 0.5$ by magenta plus signs.  The brown solid line in the  lower panel is derived from the 4 Myr PARSEC isochrone for a primary  component of 0.6 \msun mass.
\label{fig:plotcomp} }
\end{figure}
% comp => plotcomp

\begin{deluxetable}{c c c l}
\tablecaption{Meaning of the $p_{\rm phys}$ flag
\label{tab:phys} }
\tablewidth{0pt}                                   
\tablehead{   
  $p_{\rm phys}$ & $p_{\rm astro}$ & $N$ & Comment }
\startdata
0    & 0     & 397 & Astrometric non-members \\
0.1  & 0.5   & 27 & Photometric non-members \\
0.2  & 1    & 13 & $\rho > 7''$ photometric non-members  \\
0.3  & 0.5/1 & 8 &  $\Delta J < -0.5$ mag, $\rho > 7''$ \\
0.8  & 0.5   & 5 & $\rho < 2\arcsec$, $\Delta J < 2$ mag, no parallax \\
0.9  & 0.5   & 46 & Photometric members \\
1    & 1     & 91 &    Astrometric members    
\enddata
\end{deluxetable}

Figure~\ref{fig:plotcomp} plots  parameters of the  companions divided by the $p_{\rm astro}$ flag  into three groups: physical, optical, and uncertain. The upper panel shows the expected behavior, where physical pairs concentrate  at small separations and small  $\Delta J$, optical pairs show  the opposite  trend, and most  uncertain pairs  are close, lacking  Gaia astrometry for this reason.  Note there are several wide ($\rho >  7''$) pairs in which the secondary components are brighter than the primary,  $\Delta J < 0$. Some of those companions are astrometrically confirmed members  of Ori OB1, meaning that these CVSO targets are the secondary  components to brighter stars. Such pairs should not be considered in the  statistics. However, in pairs of comparable stars it is difficult  to distinguish primary and secondary components, especially  considering their variability. We include wide pairs with $\Delta J  > -0.5$ mag in our statistics and reject 8 wide pairs with brighter companions. 

A physical  companion is  expected to have  a lower temperature  and a redder  color, compared to  the main  target.  In  the lower  panel of Figure~\ref{fig:plotcomp} this trend $\Delta  J \approx 0.75 \Delta G$ (dashed line) is confirmed.    However, the  empirical slope  of 0.75,  chosen to match the trend,   is  steeper than the  slope deduced from the isochrones;  in other words,  secondary components are slightly bluer than predicted. This trend matches the CMD in Figure~\ref{fig:cmd} where most low-mass stars are located to  the left of  the isochrones, i.e. have bluer colors. This suggests a potential bias in  masses and mass ratios derived from the isochrones.   

On the other hand, most  optical pairs (blue crosses) are elevated above the  $\Delta  J = 0.75 \Delta G$ line by at least 0.6 mag and satisfy the condition

\begin{equation}
\Delta J > 0.6 +  0.75 \Delta G 
\label{eq:opt}
\end{equation}
(dotted line  in the Figure);  most optical companions are   bluer  than  the  astrometrically  confirmed  physical companions.   This allows us  to classify  the pairs that  lack good astrometry.   We set $p_{\rm phys} = 0.9$ for 46 pairs below the dotted line (\ref{eq:opt}) and $p_{\rm  phys} =  0.1$ for 27  pairs above  this line.  The close and faint companion to target 1281 (Figure~\ref{fig:triples}) is just above the line and, although possibly physical, it got $p_{\rm  phys} =  0.1$. Of  the 36 pairs  with $\rho  < 2''$  lacking  reliable GDR2  astrometry, 30  are physical and one optical  according to the photometric criterion; five pairs with  $\Delta J < 2  $ mag that lack  Gaia photometry are considered likely physical  and assigned $p_{\rm phys}  = 0.8$, based  on the low probability  of chance projections  at close  separations.   For a typical $J=13$ mag target, we  expect to find 2.7 optical companions   within  2\arcsec ~with  $\Delta  J < 2$   mag based  on the  stellar   density  in the  VISTA Orion catalog,  so one  or two  pairs  with $p_{\rm   phys}=0.8$  can still  be chance alignments. 

Selection of physical  companions based  on the  GDR2 astrometry  uses  rather loose  tolerances on  the  PM and  parallax adopted  to accommodate  effects  of unresolved  close  binaries.  One notes  in  Figure~\ref{fig:pm-plx}  blue  and  green  points (stars  with  good  astrometry) within our selection box but outside the main cluster of  points. The astrometric  filter alone does not reject all wide ($\rho  > 7''$)  optical companions, and we note in Figure~\ref{fig:plotcomp}  several  red squares  above  the  dotted line, i.e. photometric non-members; 13  such pairs  are  assigned  $p_{\rm  phys}  =  0.2$ despite  their  $p_{\rm  astro}=1$   flag.  Individual  examination of their GDR2  astrometry indeed shows  that  the  parallaxes and/or  PMs  of  both  components disagree   significantly,  although  both  satisfy  our  loose  astrometric membership  criteria.  Some remaining  wide  pairs with  $p_{\rm astro}=  p_{\rm   phys} =1$ may  still be optical, and we address  this issue in the  following statistical analysis.

The meaning of the $p_{\rm phys}$ flag and the number of companions in each group  are summarized in  Table~\ref{tab:phys}. Overall,   we consider 142 pairs  with $p_{\rm phys} > 0.5$  as physical (binaries); 135 of  those have main targets that  are deemed to be  members of Ori OB1.  The remaining pairs  either are optical (chance projections) or  do not belong to Ori OB1.

At  wide separations, the contamination  by unrelated association  members and  other stars that  pass the astrometric  and photometric  selection  criteria is  far from  being negligible.  The  density of  these  contaminants,  or  interlopers,  must be  estimated  to  make  appropriate correction  to the binary statistics. The  lower limit of  55 stars  per square degree is  obtained by star counts  in the CVSO  catalog  in section~\ref{sec:cluster}.  The upper  limit  of $\sim$4000  stars per  square degree results from  the star count  in the  Vista Orion catalog. We estimated the realistic  density of interlopers by selecting companions in the GDR2 with  separations from 20\arcsec ~to 30\arcsec ~from our targets that pass  the adopted astrometric criteria, ignoring their $r_\varpi$. The  $J$-band magnitudes of  these companions were retrieved from the  VISTA Orion catalog by positional match (typically within  0\farcs06). The resulting list of 71 companions was further filtered  by colors using equation~\ref{eq:opt}, leaving 29 companions with  $\Delta J < 3$ mag,  24 of those having  $\Delta J < 2$ mag. Assuming that all those wide companions are interlopers, we derive their density of 226(187)   stars per square degree for $\Delta J <3(2)$, respectively. The statistical error of  this estimate is about 20\%. Considering other uncertainties, we assign  the relative error of 30\% to the estimated density of interlopers.  

%-------------------------------------------------
\subsection{List of binaries in Ori OB1}
\label{sec:tabcomp}

%\input{tabcomp_small.tex}
% portion of the main table tabsample, see tabsample.pro
\begin{deluxetable*}{rr  rr rr cc}
\tabletypesize{\scriptsize}     
\tablecaption{Companions with $\Delta J < 3$ mag (fragment) 
\label{tab:comp}  }
\tablewidth{0pt}                                   
\tablehead{                                                                     
\colhead{$N$} & 
\colhead{CVSO} & 
\colhead{$\theta$} & 
\colhead{$\rho$} & 
\colhead{$\Delta J$} & 
\colhead{$\Delta G$} & 
\colhead{$p_{\rm astro}$} & 
\colhead{$p_{\rm phys}$} \\
& & 
\colhead{(degr)} &
\colhead{(\arcsec)} &
\colhead{(mag)} &
\colhead{(mag)} &
}
\startdata
   5 & 427 &304.3 &16.730 &$-$0.41 &$-$2.41 & 0.0 & 0.0\\
   6 & 432 &101.9 & 3.440 & 1.89 & 2.53 & 1.0 & 1.0\\
   8 & 438 &328.9 &15.482 & 2.08 & 0.45 & 0.0 & 0.0\\
   9 & 439 &123.5 &17.400 & 2.16 & 3.05 & 0.0 & 0.0\\
  10 & 444 &303.1 &15.419 & 1.52 & 0.78 & 0.0 & 0.0\\
  11 & 449 & 11.3 &17.845 & 2.48 & 1.00 & 0.5 & 0.1\\
  13 & 453 &100.5 &16.993 & 2.94 & 1.39 & 0.0 & 0.0
\enddata 
\end{deluxetable*}

\begin{figure}
\epsscale{1.1}
%\plotone{deltam.eps}
\plotone{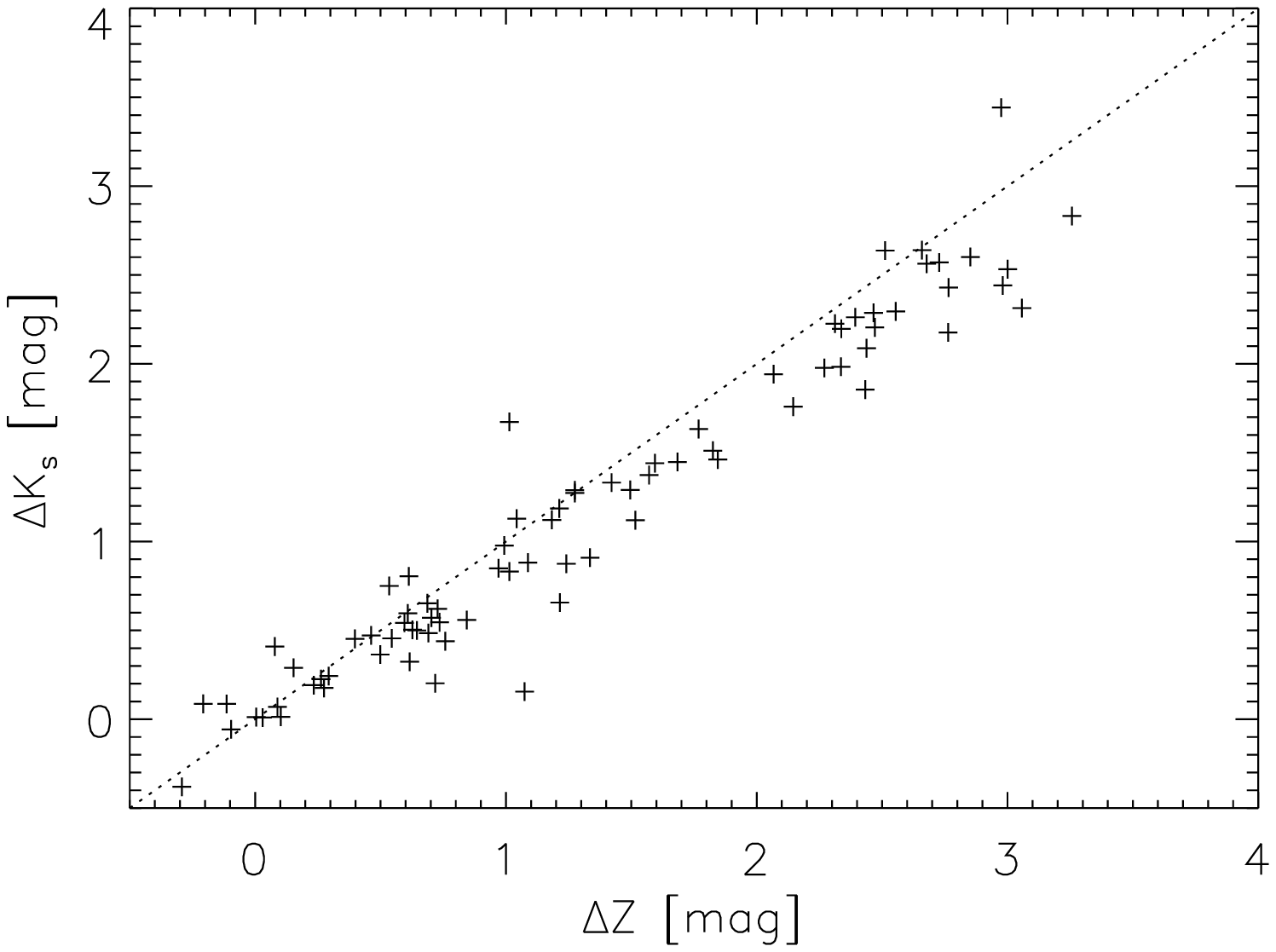}
\plotone{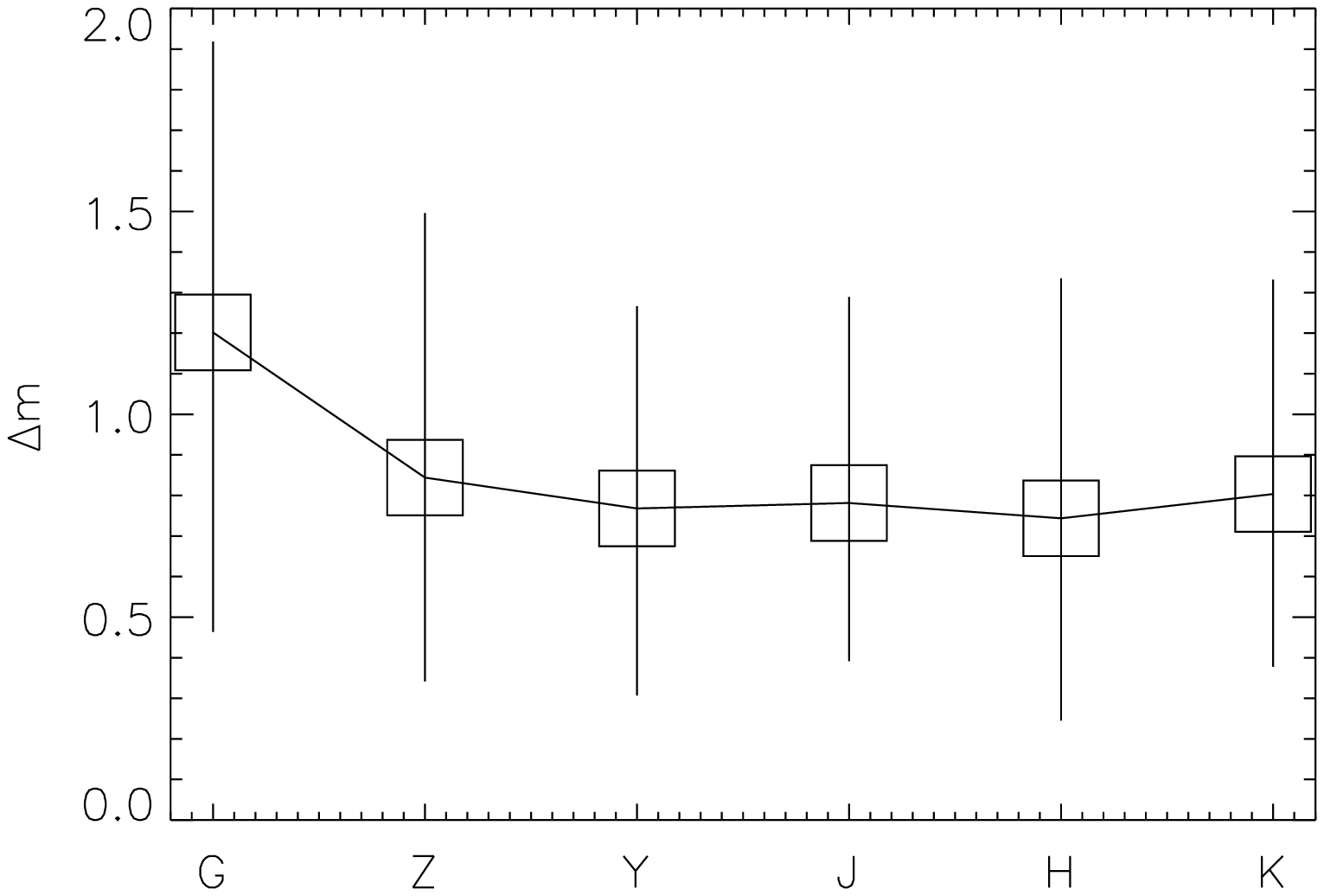}
\caption{Top: comparison of magnitude differences in the $Z$ and    $K_s$ bands for 86 physical companions with $\rho > 2''$; the    dotted line marks equality. Bottom:  Magnitude difference vs. wavelength for physical  binaries. Squares plot the median values in each band, the bars show  the first and third quartiles. 
\label{fig:deltam} }
\end{figure}

Multi-band  photometry   provided  by  the  VISTA   Orion  survey  can potentially   help    in   distinguishing   physical    and   optical pairs.  However,  magnitudes of a given low-mass young star in all VISTA nIR bands are very similar.  The top panel of  Figure~\ref{fig:deltam} compares magnitude differences of physical  binaries at the shortest and longest VISTA wavelengths. The slope is barely different from one,  while the large scatter results from a combination of variability, photometric errors, and/or circumstellar extinction.  We  base our  statistical analysis  on the magnitude differences in the $J$ band, which are very close to $\Delta m$  in the  adjacent  photometric  bands $Y$  and  $H$.  For  physical companions, the  rms differences  between $\Delta m$  in the  band $J$ compared  to the bands  $Z, Y, H, K_s$  are  0.38, 0.28,  0.29,  and 0.19  mag, respectively. Hence,  in the following table we replace $\Delta J$   by $\Delta J'$ --- the median $\Delta m$ in the $Y, J,  H$ bands --- in order to  reduce random errors and the impact of variability and taking advantage of the fact that there is no systematic difference between $\Delta m $ in these three bands.   We use $\Delta J'$ in the following in place of $\Delta J$.

Figure~\ref{fig:deltam} illustrates the distributions of $\Delta m$ in six photometric bands for physical  binaries by plotting  medians and quartiles of the distributions.  The distributions in all nIR bands are remarkably similar,  with median $\Delta  m$ around 0.8  mag. Remember that we  keep only pairs  with $\Delta J  < 3$ mag.  If  the magnitude differences were  distributed uniformly, the median would  be close to 1.5 mag; instead,  real binaries prefer small $\Delta  m$.  The median $\Delta  G =  1.2$  mag  is larger  than  in the  nIR,  as expected  for low-mass,  low-temperature companions. 

The list of  587 companions with $\Delta J < 3$ mag found in the images and including wider  pairs added from the VISTA  Orion catalog is presented in Table~\ref{tab:comp}. Its first four columns are identical to those of  Table~\ref{tab:bin}.   The  next  two columns  give  the  magnitude difference $\Delta J'$ (median over $Y, J, H$ bands) and the Gaia $\Delta G$. The last columns  contain the flags $p_{\rm  astro}$ and $p_{\rm  phys}$  introduced  above.  There are  142 likely physical pairs with $p_{\rm phys} > 0.5$.   To  distinguish  the   7 physical pairs  where the  main  star is  not  a  member of Ori OB1  (e.g. CVSO~569, see above), their $p_{\rm phys}$ are listed as negative. 

%-------------------------------------------------
\subsection{Close binaries observed at SOAR}
\label{sec:SOAR}

\begin{figure}
\epsscale{1.1}
\plotone{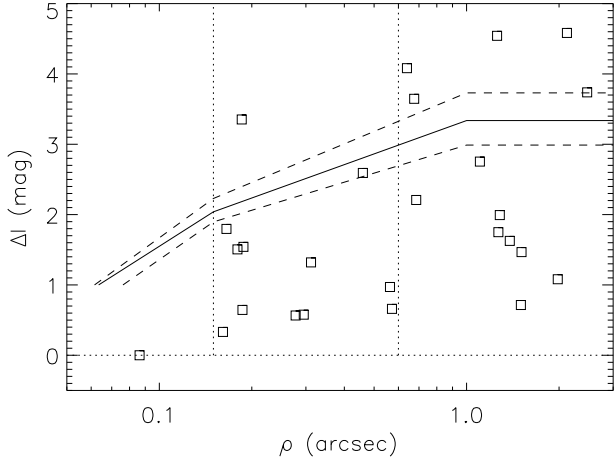}
\caption{Separations  $\rho$ and magnitude  differences $\Delta  I$ of  CVSO  pairs resolved at  SOAR (note: 2 pairs with respective separation of 3\farcs7 and 3\farcs1 are outside the plotting area). The  full and  dashed lines  show the  median  detection  limit and  its  quartiles.  The separation  range
  between 0\farcs15  and 0\farcs6 used in the  statistical analysis is  delimited by the vertical dotted lines.
\label{fig:speckle}
}
\end{figure}

To  probe binary  frequency at  smaller separations,  we  observed 123 relatively bright stars in Ori OB1 selected from the CVSO catalog with the speckle camera at the  4.1 m Southern Astrophysical Research (SOAR) telescope in 2016 January. The instrument, data reduction, and results are published in \cite{SAM19}.   We detected 28 pairs, including some wide     ones    also    found     in    the     VISTA Orion    images (Figure~\ref{fig:speckle}). The  closest pair  has  a separation  of 0\farcs09; 12  pairs have separations between  0\farcs15 and 0\farcs6, and  all  these  detections  are  reliable.   Some  close  CVSO  pairs discovered in  2016 were confirmed by further  speckle observations in 2017--2019.  The SOAR data are  published and we do not duplicate them here.

Only   74  stars   observed   at  SOAR   overlap   with  the   present CVSO-VISTA-Gaia  sample, the  rest are located  outside  the sky region studied  here.  We can use the higher spatial resolution of SOAR observations to double-check VISTA detection sensitivity at small  separations, although restricted to the overlap sample. The SOAR observations confirmed that no companions with separations $>$1\arcsec ~were missed in the VISTA companion search, thereby affirming the assumed  completeness for  separations $>$1\farcs2 and $\Delta  J < 3$ mag. At smaller separations, i.e.\ 0\farcs6--1\farcs2, there is only one binary resolved at SOAR that was missed by VISTA (No. 214, CVSO~2001, 0\farcs64, $\Delta I = 4.1$ mag,  close to or below the VISTA detection limit). 

Since this paper is devoted to wide  binaries in our well-defined sample, we evoke the SOAR data on  closer pairs in a {\em different}  sample only for reference.  The  SOAR sample  is, on average,  brighter than the  CVSO-VISTA-Gaia sample.   More massive stars are expected to have  an increased binary frequency and, indeed, the  data hint  at a  larger companion  fraction in  the  SOAR sample, although the difference is not statistically significant.  Moreover, a higher companion fraction among the SOAR sample is naturally expected  due  to the general shape of the separation distribution that has a peak at $<$100\,au.

%---------------------------------------------------------
\section{Binary statistics}
\label{sec:stat}

In this section,  we study the statistics of  real (physical) binaries with separations  from 0\farcs6 to 20\arcsec  ~and $\Delta J  < 3$ mag discovered in  the VISTA  images and in  the VISTA Orion  catalog. The parent  sample  of 1021  members  of  Ori  OB1 contains   135  physical companions, including 4 triples;  131 members have at least one companion.   We ignore wide secondary companions that are brighter than  our targets by more than 0.5 mag in the $J$ band. 

%---------------------------------------------------------
\subsection{Distribution of the mass ratio}
\label{sec:q}

\begin{figure}
\epsscale{1.1}
\plotone{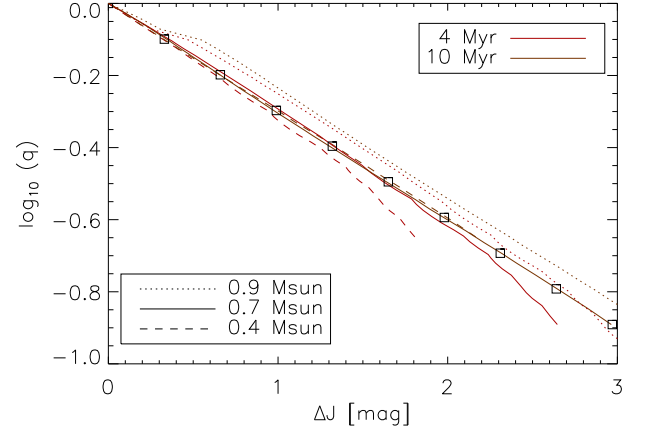}
\caption{Relation between $\log_{10} q$ and $\Delta J$  derived from  the PARSEC isochrones for primary stars of 0.4, 0.7, and 0.9 \msun  mass  and companions more massive than 0.075 \msun. Squares correspond to the linear formula $\log_{10} q = -0.3 \Delta J$.
\label{fig:massratio} }
\end{figure}

\begin{figure}
\epsscale{1.1}
\plotone{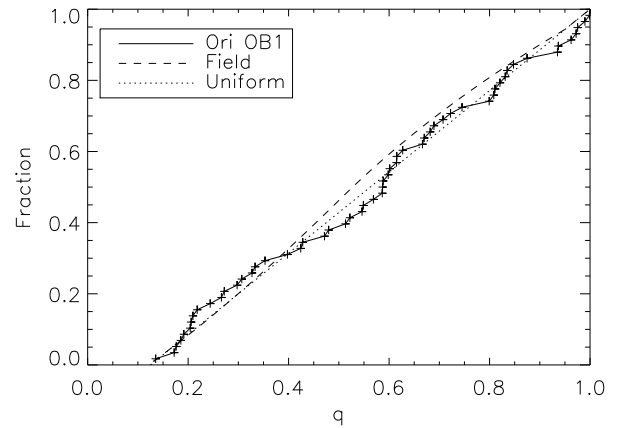}
\caption{Cumulative distribution of the mass ratio $q$ for 58 binaries  with   separations  between   1\farcs2  ~and   5\arcsec   (line  and  crosses).  The  dashed and  dotted  lines  correspond  to the  field  binaries (see the text) and to the uniform distribution, respectively. \label{fig:qhist} }
\end{figure}

As noted above, we do not attempt to derive  masses and  mass ratios. The empirical distribution of mass ratios is evaluated  here only to access the fraction of missed companions at close  separations and to quantify the survey depth. Fortunately, this fraction and the associated incompleteness correction are small and hence have little influence  on our results. 

First, we  study the distribution of the  magnitude difference $\Delta J$ for 57 binaries  in the 1\farcs2--5\arcsec ~separation range, where the detection is complete to $\Delta  J < 3$ mag and the contamination by   random  pairs   of  association   members  is   negligible.   The distribution is non-uniform, with a preference of small $\Delta J$, as can  be  inferred   by  looking  at  Figure~\ref{fig:plotcomp}.   Only 10/58=0.17 fraction of pairs have $2  < \Delta J <3$ mag, so by further restricting  our  analysis to  $\Delta  J <  2$  mag  we would  reject $\sim$17\% of binaries.

The  PMS  stars  and  their  companions in  Orion  OB1  are  variable. Therefore, there is no  unique relation between  magnitude difference and mass ratio $q$; for any  given binary, different values of $q$ can be  inferred from observations  at different  moments or  in different photometric bands.   Different circumstellar extinction (so-called infra-red companions) and  accretion rates can further complicate  translation of relative photometry into $q$.  Here we adopt the relation $ q \approx 10^{-0.3  \Delta  J}  $ derived  from  the  4  and 10  Myr  isochrones (Figure~\ref{fig:massratio}).   There  is  little  dependence  on  the age, except maybe for the youngest and lowest mass stars in Ori OB1b, which represent at most $\sim$15\% of the binaries. Our formula is an  excellent approximation for 0.7 \msun ~primary components,  roughly equivalent to a spectral type K7-M0, while for 0.9 \msun ~stars it works less well, with an rms error of 0.06  dex in $\log_{10} q$.  For 0.4 \msun ~stars at 4\,Myr the derived mass ratios could be overestimated by  0.1 dex in $\log_{10} q$, but  the number of binaries in this parameter space is small; their hydrogen-burning companions have $\Delta J < 2$ mag.  
%A variability by  0.5 mag in $J$ would  cause a $q$  error of  0.15  dex, typically larger than the  errors of  the formula or the age dependence,  making our crude  approximation  acceptable.

The mass ratios deduced by the crude formula from $\Delta J$ appear to be  distributed  uniformly  (Figure~\ref{fig:qhist}). Reassuringly,  a uniform  distribution of  $q$ is  known to  hold for  field solar-type binaries \citep{R10,FG67}.  \citet{El-Badry2019} studied the mass ratio distribution of wide field binaries,  grouping them  by the  primary mass  and  separation.  They modeled the distribution by a broken power law plus some twins with $q > 0.95$.  We average the model parameters  from their Table~F1 in the mass range from 0.4 to 0.8 \msun ~and the separation range from 600 to 2500 au,  roughly matching  our  survey, and  adopt  $\gamma_{\rm smallq}  = 0.25$, $\gamma_{\rm  largeq} = -0.8$,  and $f_{\rm twin} =  0.02$. The corresponding    cumulative    distribution    is    over-plotted    in Figure~\ref{fig:qhist} by the dashed  line; it barely differs from the uniform distribution.

Adopting a uniform distribution of $q$, we evaluate the fraction of missing binaries with small separations using the detection limit $\Delta J (\rho)$ in equation (\ref{eq:detvista}). This  correction, relevant only at separations below 1\farcs2, is small (factor $\sim$1.12, see below). Therefore, our  results  are  little   affected  by  the assumptions  regarding the  mass  ratio conversion  and the  detection limit.

\begin{deluxetable*}{ c c    cccc  ccc  ccc }
\tabletypesize{\scriptsize}     
\tablecaption{Multiplicity on Ori OB1 
\label{tab:sep}  }
\tablewidth{0pt}                                   
\tablehead{                                                                     
\multicolumn{2}{c}{Separation} & \multicolumn{4}{c}{Full sample} & \multicolumn{3}{c}{Ori OB1a} &  \multicolumn{3}{c}{Ori OB1b}  \\
$''$ & au   & $N_i$ & $N'_i$ & $N_{\rm rand}$ & $f_i$  & $N_i$ & $N'_i$ & $f_i$  & $N_i$ & $N'_i$ & $f_i$ 
}
\startdata
0.6 -- 1.2  & 314  & 26 & 24 & 0.1 &  9.4$\pm$1.9 & 16 & 16 & 9.0$\pm$2.3 & 10 &  8 &10.2$\pm$3.4 \\
1.2 -- 2.4  & 627  & 33 & 28 & 0.2 & 10.7$\pm$1.9 & 16 & 13 & 8.0$\pm$2.1 & 17 & 15 &15.5$\pm$3.9 \\
2.4 -- 4.8  & 1255 & 23 & 19 & 1.0 & 7.2$\pm$1.6 & 11 & 10 & 5.2$\pm$1.8 & 12 &  9 &10.7$\pm$3.3  \\
4.8 -- 9.6  & 2511 & 16 & 16 & 3.9 & 3.9$\pm$1.4 & 10 & 10 & 3.8$\pm$1.7 &  6 &  6 & 4.2$\pm$2.5  \\
9.6 -- 19.2 & 5023 & 31  & 18& 15.5& 5.1$\pm$2.4 & 20 & 10 & 5.1$\pm$2.8 & 11 &  8 & 5.0$\pm$3.5  \\
%\hline
0.6 -- 9.6 &\ldots& 98 & 87 & 5.2 &  9.4$\pm$0.9 & 53 & 49 & 7.8$\pm$1.1 & 45 & 38 &12.2$\pm$1.8  
\enddata 
\end{deluxetable*}

%-------------------------------------------------
\subsection{Separation distribution and companion fraction}
\label{sec:sephist}

\begin{figure}
\epsscale{1.1}
\plotone{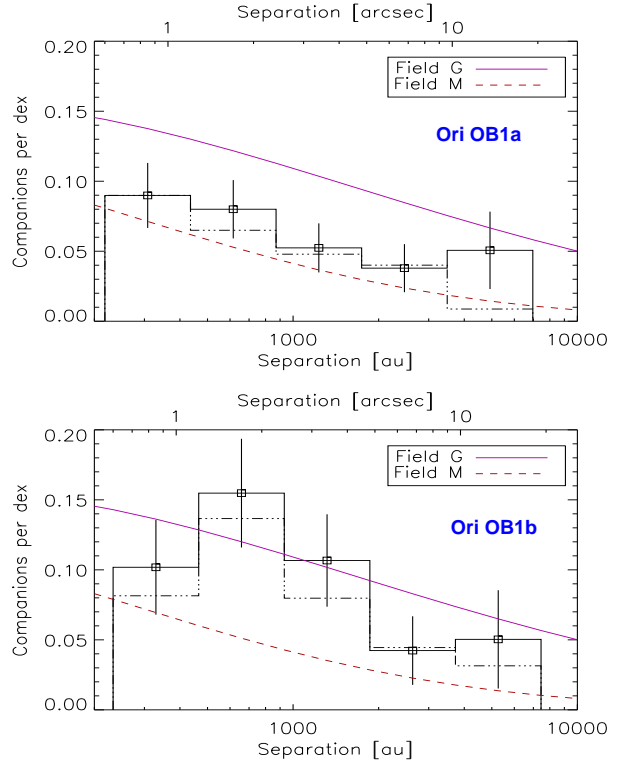}
%\plotone{sephist-b.eps}
\caption{Companion frequency  vs. separation (fraction  per decade) in  Ori OB1a (top)  and Ori OB1b (bottom). The  full line corresponds to  binaries with $\Delta J < 3$ mag, the dash-dot line to binaries with  $\Delta J  < 2$ mag.   The log-normal distributions for  field stars  (solid and dashed curves) are described in the text.
\label{fig:sephist} }
\end{figure}

The  distribution  of  separations  and the  companion  frequency  are determined in  five logarithmic bins  of 2$\times$ width  covering the separation range from  0\farcs6 to 19\farcs2 (1.5 dex,  222 to 7104 au at 370\,pc distance).  Binaries with $\Delta  J < 3$ mag and $\Delta J < 2$  mag are  counted in each  bin. These numbers $N_i$ are  used to compute the companion frequency per decade of separation $f_i = (N_i - N_{\rm rand}) / (N_{\rm tot} \log_{10} 2)$, where $N_{\rm tot}$ is the sample  size.  The errors of $f_i$ assume the Poisson statistics.  The first  bin is  corrected for  undetected companions relatively to other bins; however,  this correction is minor, 1.12 and 1.04 times for the two  $\Delta J$ thresholds.    The expected number of random pairs $N_{\rm rand}$ is estimated from the  density  of  potential  contaminants (interlopers)  deduced  in  section~\ref{sec:phys}    by    selecting    companions    in    the  20\arcsec-30\arcsec   ~separation  range   and  applying   the  same astrometric  and   photometric  filters  as  used   for  the  closer  companions: 226 and  187 stars per square degree for  $\Delta J < 3$  mag and $\Delta J < 2$  mag, respectively; a relative error of 30\% is  assumed for the density of interlopers and included in the estimated errors of $f_i$.  

The separation distribution  for the full sample and  for the OB1a and OB1b groups  is given in  Table~\ref{tab:sep}, where $N_i$  and $N'_i$ are the numbers  of pairs with $\Delta J  < 3$ mag and $\Delta  J < 2$ mag, respectively, in each bin  and $f_i$ are the fractions per decade for $\Delta  J < 3$ mag in  per cent, corrected for  incompleteness in the first bin  and for  the random  pairs.  The estimated numbers $N_{\rm rand}$  of  contaminants  with $\Delta J < 3$  mag are listed for  the  full sample.  They become  comparable to  the actual  number of  companions in the last  bin, increasing the $f_i$ error. Considering  this uncertainty, the  last line gives the total  number of binaries and the companion  frequency  in  the first four bins only covering 1.2 dex in separation. 

The separation distributions  in Ori OB1a and Ori  OB1b are plotted in Figure~\ref{fig:sephist}.  Projected separations are translated into au using distances of 363 and 388 pc, respectively  (section~\ref{sec:main});  a common distance of 370 pc is used for the full sample.  The distributions  appear to be  different, especially in the second  bin,  where there are  almost twice as many binaries in Ori OB1b as in Ori OB1a. However, bear  in mind that the  projected separation $s$ equals the  semimajor axis $a$ only statistically  and their ratio $s/a$ varies by a factor of 2  (i.e. the bin width) both ways owing to projections  and  random  orbital  phases.   Therefore,  even  if  the distribution of semimajor axes had a sharp feature, it would be spread over adjacent bins in the distribution of $s$.   Modeling  shows  that the distribution  of $s/a$  depends on  the eccentricity  distribution:  its median is  0.9 when  the average  eccentricity is  around  0.5 and  0.98  if the  eccentricity  distribution is  linear  (thermal), $f(e) = 2e$  (see the Appendix). The  latter is appropriate for wide binaries  considered here \citep{Tok2020}. Although various correction factors  on the  order of 1 have  been proposed in the  literature to convert  $s$ into $a$,  no correction is actually needed  and the statistical  distributions of  $\log s $ and $  \log a$ can  be compared directly,  provided they are smooth on a $>$0.3 dex scale. 

The different  multiplicity fractions in  both groups can be  seen even from the raw  numbers.  In Ori OB1a, we detect  74  binaries out of 658 targets (multiplicity 11.2$\pm$1.3\%).  In  contrast, in Ori OB1b there are 61 binaries among  363 targets (multiplicity 16.8$\pm$2.2\%).  The difference of  5.6$\pm$3.2\% is significant at  the $1.8\sigma$ level. The   corrected   multiplicity  fractions   in   the   last  line   of Table~\ref{tab:sep}  differ by  4.4$\pm$2.1\%, their ratio is  1.6$\pm$0.3. 

For  reference,  we plot  in  Figure~\ref{fig:sephist} the log-normal distribution for field solar-type binaries derived by \citet{R10} with a median  of 50\,au,  separation dispersion of  1.52 dex (2.28  dex in period),  and companion  frequency of  0.60.  The  dotted line  is the log-normal   distribution   for   field   M-type   dwarfs   found   by \citet{Winters2019}, with  the median at 20\,au,  dispersion 1.16 dex, and the companion frequency of 0.35, appropriate for early M dwarfs. While the separation distribution of binaries in Ori OB1a is consistent  with those of early M-dwarfs, as expected given that the distribution of spectral types in the CVSO sample is weighted to M stars, Ori OB1b shows a high binary fraction, that partially is even higher than for solar-type dwarfs. In both sub-groups of Ori OB1, the companion frequency at separations above $\sim$4000\,au (in  the last bin) has a large uncertainty caused by the substantial and uncertain fraction of contaminants. 

As mentioned above, we refrain from assigning individual masses to the stars of our sample.  Instead, we use the $J$-band magnitude as a proxy for mass, in order to explore the dependence of multiplicity on stellar mass. We split the members of Ori OB1 into  three equal sets grouped by $J$ magnitude: brighter  than $J=12.63$,  intermediate, and  fainter  than $J=13.56$, with 340 stars in each group.  The numbers of non-single stars in these groups  (61, 48,  and 26,  respectively, or multiplicity  fractions  0.18$\pm$0.023,  0.14$\pm$0.02, and  0.08$\pm$0.016)   suggest  a substantial dependence of multiplicity on mass, as observed for field binaries.

%-------------------------------------------------
\subsection{Close binaries}
\label{sec:close}

\begin{figure}
\epsscale{1.1}
\plotone{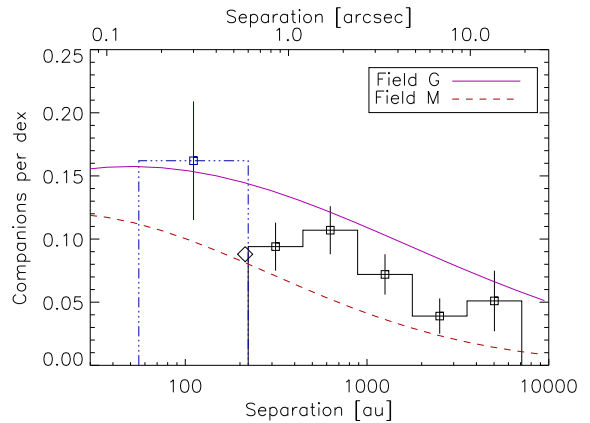}
\caption{Distribution of  separation in the full  sample. The dash-dot line  shows the  result for  the SOAR  sample.  The  diamond  is the  companion frequency in the ONC from \citet{Reipurth2007}.  A distance of 370\,pc is used.
\label{fig:close} }
\end{figure}
% sephistplot.pro 

Figure~\ref{fig:close}   presents   the   distribution  of   projected separations  in  the  full  CVSO-VISTA-Gaia  sample,  including  the  observed frequency of closer pairs derived from the SOAR observations (the wide dash-dot bar).  The latter is $12/123=0.162\pm0.047$ in the separation range  from 0\farcs15  to 0\farcs6.   The faintest  companion  in this range has a magnitude  difference $\Delta  I = 3.4$  mag. If  it is excluded,  the binary  frequency  would be  $0.148  \pm 0.045$.   SOAR observations have variable resolution and contrast sensitivity because they used a laser to  correct for atmospheric turbulence and  depended on  the variable atmospheric conditions.  The median  detection limit is $\Delta I \sim 2$ mag.  Therefore, the SOAR  survey is somewhat shallower compared to our   survey.  The   SOAR   sample, selected from amongst the brightest CVSO stars,   differs   from   the CVSO-VISTA-Gaia  sample  in that it  contains,  on average,  brighter, earlier K-type, more  massive stars, known  to have  a larger multiplicity   in comparison with late K- and M-type  dwarfs. Considering  these  differences,  direct  comparison between the SOAR and VISTA Orion multiplicity surveys  has to be done with caution. All we can affirm is a qualitative agreement.

\begin{deluxetable}{l ccc cc}
\tablecaption{Close binaries in Gaia
\label{tab:DR2} }
\tablewidth{0pt}                                   
\tablehead{   
Group  & $N_{\rm tot}$ & No $\varpi$ & $r_\varpi > 2$  &   $f_{\rm close}$ }
\startdata
OB1a  & 658       & 25   & 65  & 0.137$\pm$0.014 \\
OB1b  & 363       & 27   & 35  & 0.171$\pm$0.022 
\enddata
%\footnote{No.\ stars without measured  parallaxes}
\end{deluxetable}

Yet another way to probe the frequency of close binaries is offered by Gaia. Stars  without measured  parallaxes are  almost certainly close binaries  with separations  between  0\farcs1 and  0\farcs7,  as  demonstrated, e.g., by \citet{TokBri2019}.  Moreover, stars with  excess parallax error are also likely close binaries.  The total  number of these close binary  candidates  can be  used to  estimate the  frequency of  close binaries $f_{\rm close}$, with  the caveat that the exact  range of separations and mass ratios of these close binaries is not defined and the numbers are  not directly  comparable  to the  frequency  per decade  computed above.  The  numbers are reported in Table~\ref{tab:DR2}.  We note the increased  fraction of   candidate  close binaries  in  Ori OB1b  compared to  Ori OB1a. This echoes the difference between these groups found for wider pairs,  although the difference  between the  frequency of  close  Gaia binaries, 3.6$\pm$2.6\%, is not statistically significant.

\citet{Reipurth2007} measured the companion frequency for 781 low-mass stars  in the periphery  of the  ONC  at projected separations  from 67.5  to 675  au and  found it  to  be 8.8$\pm$1.1\% (represented by the diamond symbol in Figure~\ref{fig:close}).  Their result agrees, within errors, with the multiplicity  in Ori OB1. However, the decline in the  binary frequency at separations beyond  $\sim$200 au suggested by their study is certainly refuted by our data. Even in the ONC, \citet{Jerabkova2019} found a substantial number of binaries with separations from 1 to 3 kau.  The 14  low-mass  binaries in the  ONC with  separations from  30 to  160 au  discovered  by  \citet{defurio2019} match, within errors,  the  frequency of  M-type  binaries in the field. 

%---------------------------------------------------------
\subsection{Spatial distribution of binaries}
\label{sec:skybin}

\begin{figure}
\epsscale{1.1}
\plotone{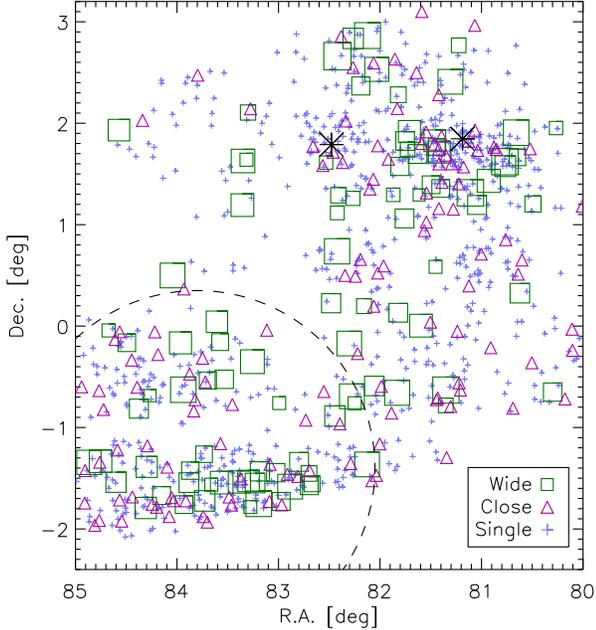}
\caption{Spatial  distribution of single  stars (blue  crosses), close Gaia  binaries (magenta  triangles), and wider  VISTA Orion binaries  (green squares).  The size  of the squares reflects binary separation ranging from  0\farcs6 to 10\arcsec.   The two black  asterisks show the locations of 25~Ori and HR~1833. The dashed circle depicts the  boundary of Ori OB1b.
\label{fig:skybin} }
\end{figure}
% plotsample.pro => plotskybin

The difference in  the binary statistics between the  two subgroups of Orion  OB1  is  intriguing.   To   clarify  it  further,  we  plot  in Figure~\ref{fig:skybin}    the   spatial   distribution    of   single (i.e.  unresolved) stars,  wide binaries  with separations  between  0\farcs6   and  10\arcsec,  and   potential close  binaries inferred from  Gaia.  Binaries  with separations $>$10\arcsec ~are ignored  because several of them are likely random pairs  of   association  members.  

\begin{figure}
\epsscale{1.1}
\plotone{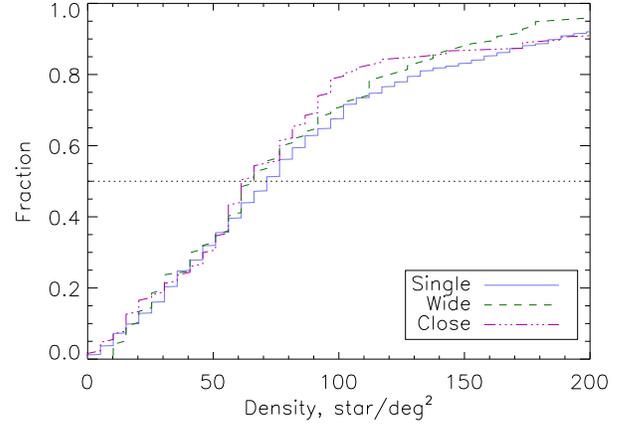}
\caption{Cumulative distributions of the density of association members around single stars, wide, and close binaries, computed for 0\fdg25 radius.
\label{fig:density} }
\end{figure}
% cluster.pro => density2.eps

One notable feature of  Figure~\ref{fig:skybin} is the  apparent  absence of wide binaries in the  two stellar over-densities of Ori OB1a  near the stars 25~Ori and  HR~1833  (marked by  asterisks).   The  latter  group lacks  wide binaries completely. In the  25~Ori cluster, wide binaries are located on the periphery, at a distance of $\sim$0\fdg5 from the central star, and avoid  the center.  Meanwhile,  both clusters contain  closer   Gaia binaries.

We estimated the  mean stellar surface density around each  target by counting association  members  within 0\fdg25 radius, which gives an average number of 11 stars in this area. Figure~\ref{fig:density} plots  the cumulative distribution of surface density  in  the  three  groups:  single  stars,  wide  binaries  with separations  between  0\farcs6   and  10\arcsec,  and  close  binaries inferred from  Gaia (763,  97, and 127  stars, respectively).  The distributions do not differ significantly  (the two-sided Kolmogorov-Smirnov test gives a probability of 0.60 for wide binaries and single stars having the same parent distribution). Therefore, we cannot claim that the frequency of wide binaries  in Ori OB1 depends on the stellar density.

On the other hand, the region has dynamically evolved and the density structures seen today most likely differ from the actual birth configuration. Yet, stars in Ori OB1a appear more clustered than those in Ori OB1b, despite being  roughly twice the age. Adopting the velocity dispersion in a subgroup of 0.5 mas~yr$^{-1}$, which corresponds to 0.8 km~s$^{-1}$,  consistent with typical velocity dispersion of young stellar groups and clusters in Orion OB1 \citep{briceno2007,Kounkel2018}, we find that  in 7 Myr stars can move from their birthplace by $\sim$1\degr.  Stars within $\sim$0\fdg5 from the 25~Ori cluster center could have experienced dynamical interactions and been ejected  6-7~Myr ago.  This timescale is nicely consistent with the age of the 25~Ori group \citep{CVSO}. On the other hand, wide binaries around it could have formed in a low-density environment rather than in the cluster. If the 198 stars located within 0\fdg5 from 25~Ori and HR~1833 are removed from the Ori OB1a sample, the  frequency of binaries closer than 9\farcs6 among the remaining 460 members becomes  larger, $\sim$0.1, similar to their frequency in Ori OB1b. 

%---------------------------------------------------------
\section{Discussion and summary}
\label{sec:disc}

\begin{figure}
\epsscale{1.1}
\plotone{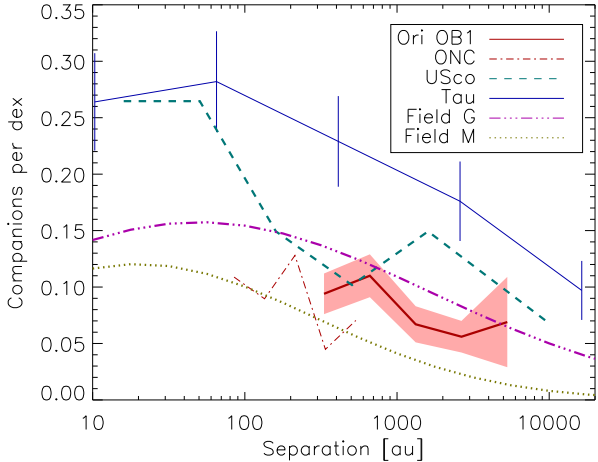}
\caption{Comparison of multiplicity in different populations (see text).
\label{fig:multcompare} }
\end{figure}

In Figure~\ref{fig:multcompare}  we put our results in  the context of multiplicity in other regions, with  the obvious caveat related to the differences  in the  stellar mass  range and  completeness  of various surveys. Assuming a uniform distribution of $q$ between 0.05 and 1, we correct  our   estimate  of  multiplicity  fraction  in   Ori  OB1  by 0.95/0.9=1.055. The band shows the $1 \sigma$ statistical errors.  The canonical companion  frequency of 0.6 for the  field solar-type dwarfs \citep{R10}  is assumed.  The  multiplicity in  the Upper  Scorpius OB association measured  by \citet{TokBri2019} is  similarly corrected by 0.95/0.7=1.36, considering  their lower  mass ratio limit  of $q > 0.3$. Their  data in  the mass  range from  0.4 to  1.5 \msun  ~are averaged because no  clear dependence  of multiplicity on  mass was  found.  We plot   the    multiplicity   in    Taurus   using   the    data   from \citet{Joncour2017},    without    any    correction.     Data    from \citet{Reipurth2007} for the ONC are also left uncorrected because the lower limit of the mass ratio in their survey is not known.

The  diversity  of  the   multiplicity  statistics  in  young  stellar populations, noted  already by  \citet{King2012} from the  scarce data available at the time, is emerging with a stronger confidence from the modern large multiplicity surveys, including this one. In the overall Orion OB1 association the binary fraction is somewhat less than in the field;  the opposite is    true in Taurus, where the excess of binary fraction over the field is well documented \citep{Duchene2013}. In the separation range of 1.2 dex from 222 to 3552 au that corresponds to the first four bins in Table~\ref{tab:sep}, the sample of 142 stars in Taurus counts 31 companions \citep{Joncour2017}, hence $f=0.22 \pm 0.04$ \citep[the same number can be deduced by summing the last 4 bins of Fig.~4 in][]{Kraus2011}. In Ori OB1, the multiplicity in the same separation range corrected by 1.055 is $0.099 \pm 0.009$, and the difference of  $0.12 \pm 0.04$ is statistically significant. \citet{Deacon2020} found a significant deficit of binaries with separations 300--3000 au in open clusters compared to the field and moving  groups, confirming the critical role of environment in the binary population statistics.  

Considering Ori OB1a and OB1b individually, we note that OB1b shows a binary fraction comparable to the field and is more similar to Upper Scorpius, at least for the wide binaries that we probe in our study. There is accumulating evidence that Upper Scorpius, as well as other OB associations, most likely were formed in a configuration similar to how they appear today --- i.e.  as an assembly of loose stellar groups with moderate to low stellar density \citep{WriMam2018,Lim2019}. If the initial stellar density of a star forming environment largely  dictates the formation or dynamical destruction of wide binaries, one might speculate that a different stellar birth density causes the observed difference in  binary fraction between Ori~OB1a and Ori~OB1b. In this picture Ori~OB1a would have  formed from a dense cluster, while Ori~OB1b would stem from a wide-spread population with only some sparsely clustered substructures. Future kinematic studies with  Gaia providing higher precision than  GDR2 will hopefully allow to further explore this issue. Regarding the origin of the field binaries,  a mix of  Orion-type  and  Taurus-type binary populations in  a  suitable proportion would resemble the field.

The main results of our study are as follows:
\begin{itemize}

\item
Double stars  in a sample  of 1021 low-mass PMS stars  of the  Orion OB1 association, selected  from  the  CVSO  catalog, have  been identified  by the analysis  of the  nIR images  from the  VISTA Orion mini-survey.   Using   Gaia  astrometry  and  photometry, we  rejected unrelated pairs and arrived at  a list of  135  most likely real (physical)  companions, arranged in 127 binaries and 4 triples, with  projected separations from 0\farcs6  to 20\arcsec (222 -- 7400 au at 370\,pc) and magnitude difference $\Delta J < 3$ mag. 

\item
The distribution of magnitude  difference $\Delta J$ of these binaries is compatible  with a uniform  mass ratio distribution. Our  survey is almost complete for wide binaries with mass ratios above 0.13.

\item
We  found that  the two  sub-groups,  Ori OB1a  and Ori  OB1b,  likely have  a different  multiplicity  rate:  0.078$\pm$0.011  and  0.122$\pm$0.018, respectively, in  the 1.2 dex  separation range from 0\farcs6 to 9\farcs6  (222 to 3600 au at 370\,pc). 

\item
The  frequency of  wide  binaries in  Ori  OB1 depends  on mass  (more companions  around more  massive  stars  similarly to the field) but  is  independent of  the currently observed surface density of stars. Location of wide binaries on the sky suggests that they avoid cluster centers.

\item
Our  survey  highlights  the  differences in  multiplicity  properties between star-forming  regions.  The binary  population in the  field could result from a mixture of these diverse populations.
\end{itemize}

\acknowledgements

Based on observations made with ESO Telescopes at the La Silla Paranal Observatory under  programme ID 60.A-9285(B). We acknowledge the great work done by the VISTA consortium who built and commissioned the VISTA telescope and camera.  This  work used bibliographic references from the  Astrophysics Data System maintained  by SAO/NASA.  We used the  data from  the European Space Agency (ESA) mission  Gaia (\url{https://www.cosmos.esa.int/gaia} processed by the Gaia Data Processing and Analysis Consortium (DPAC, \url{https://www.cosmos.esa.int/web/gaia/dpac/consortium}).    Funding for the DPAC has been provided by national institutions, in particular the institutions participating in the Gaia Multilateral Agreement.  This research has made use of the VizieR catalogue access tool, CDS, Strasbourg, France (DOI: 10.26093/cds/vizier). The original description of the VizieR service was published in A\&AS 143, 23. The work of Tokovinin and Brice\~no is supported by NOIRLab, which is managed by Association of Universities for Research in Astronomy (AURA) under cooperative agreement with the USA National Science Foundation.

\appendix
%\twocolumn

\section{Relation between projected separation and semimajor axis}
\label{sec:simsep}

\begin{figure}
\epsscale{1.0}
\plottwo{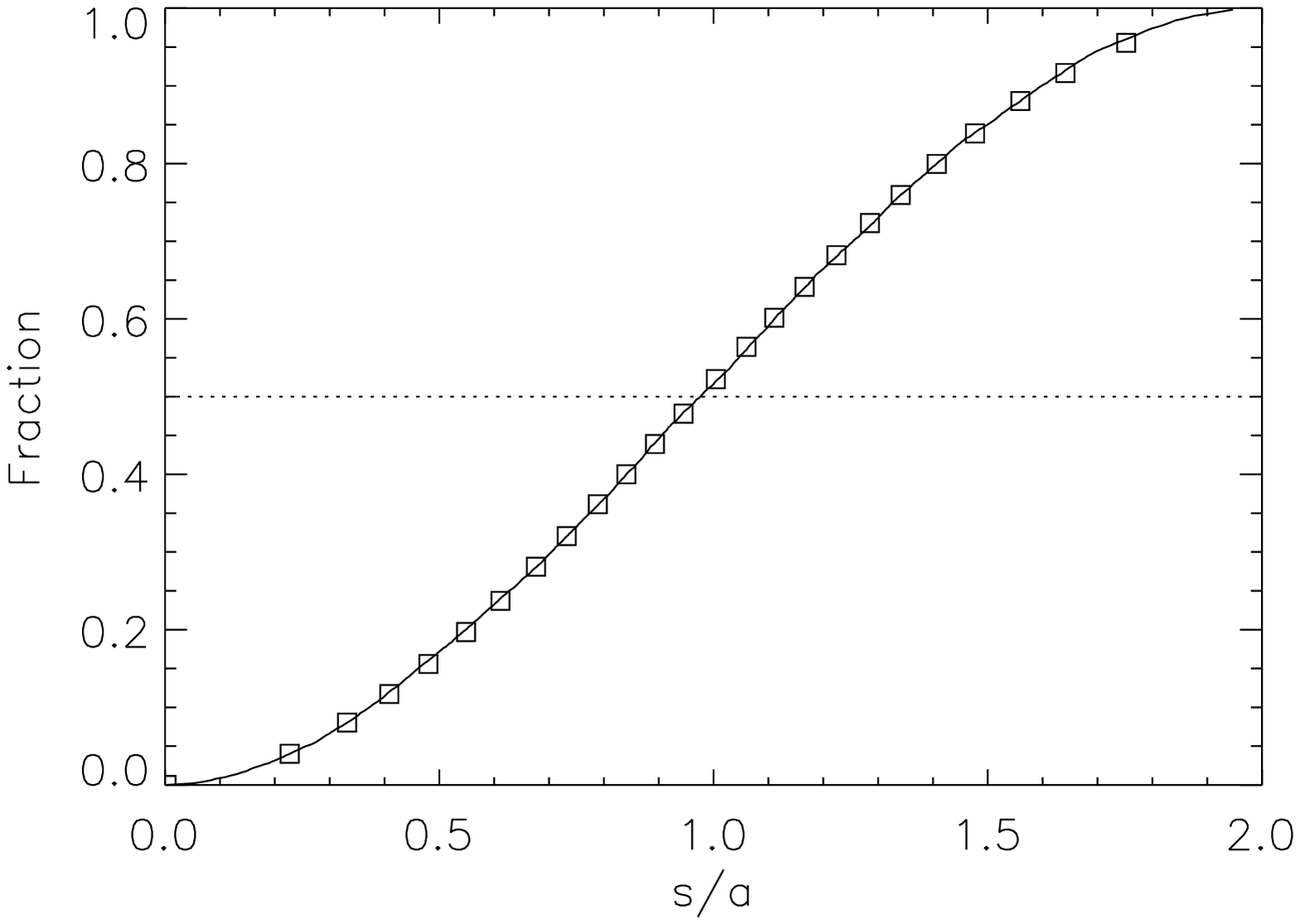}{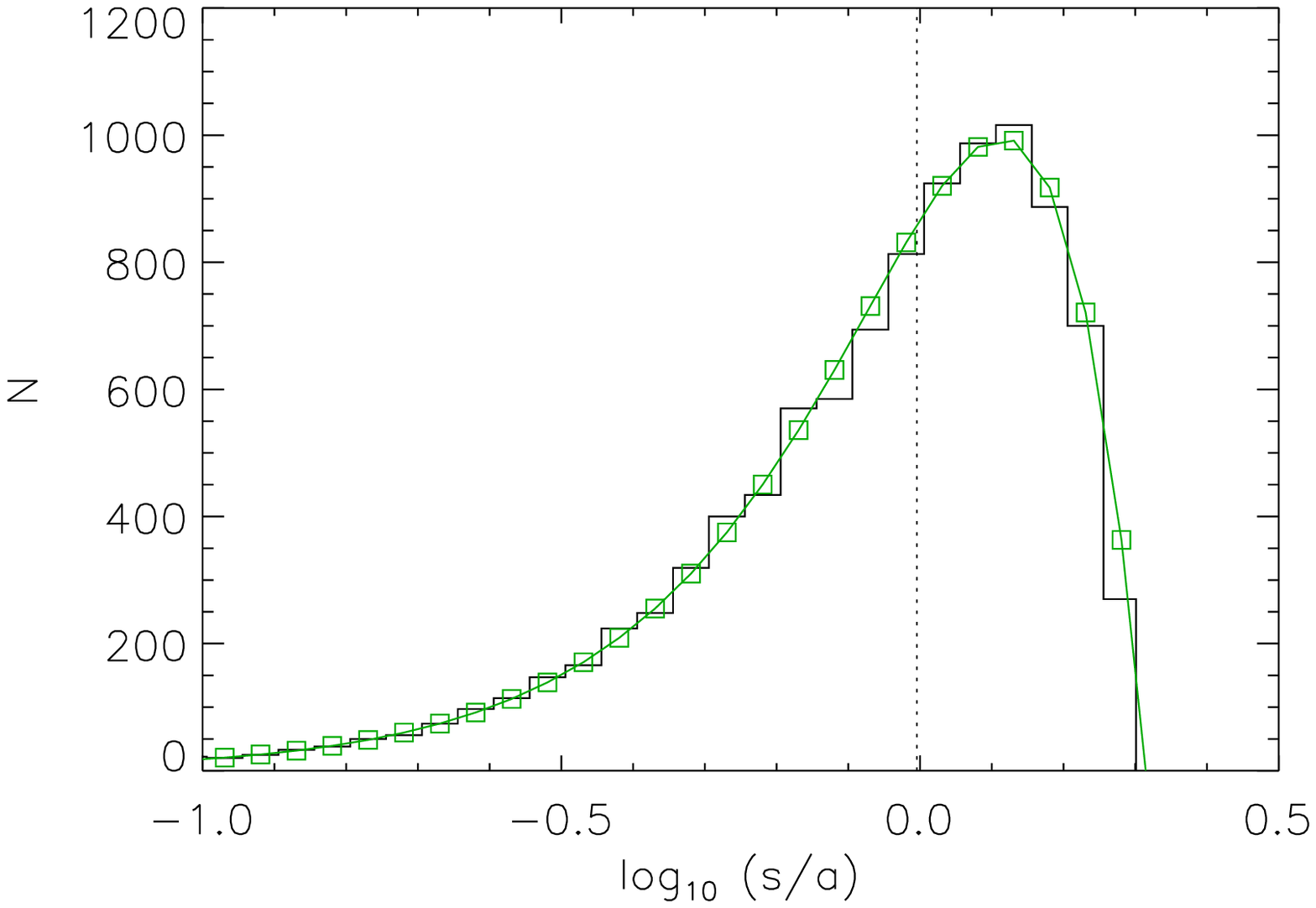}
\caption{Left: cumulative distribution of the ratio $s/a$ (line) and its analytic model (squares) for thermal eccentricity distribution. Right: histogram of the same data and their model (green squares and line); vertical dotted line marks the median.
\label{fig:simsep} }
\end{figure}
% simsep.pro => simsep.eps

On Referee's request, we include a short discussion of the statistical relation between projected separation $s$ used throughout this paper and the true semimajor axis $a$ of the binary orbit. The ratio $x = s/a$ is computed for $10^4$ simulated binaries with random orbit orientation and random phase. The resulting distribution of $x$ slightly depends on the adopted eccentricity distribution; here we assume the thermal distribution $f(e)=2e$ appropriate for wide binaries. Figure~\ref{fig:simsep} (left) shows the cumulative distribution of $x$ for simulated binaries and its analytic model 
\begin{equation}
F(x) = 0.5 + 0.5 \sin [\pi/2 \; (x/x_0)^\alpha ] ,
\label{eq:simsep}
\end{equation}
where $x_0 = 0.98$ is the median and $\alpha = 0.94$ encodes deviation from the pure sine curve. These parameters were fitted to the simulated distribution. The analytical model (\ref{eq:simsep}) is remarkably good, with rms deviation of 0.0005 and the maximum deviation of 0.005. If a uniform eccentricity distribution is assumed, the fitted parameters are $x_0 = 0.92$ and $\alpha = 1.05$, but the difference between distribution and its model is $\sim$6$\times$ larger than for thermal eccentricities. 

For a thermal eccentricity distribution, the median projected separation $s$ is an accurate measure of the true median semimajor axis,  no correction is needed. In a sample of binaries, 82.8\% of projected separations differ from the true semimajor axis by a factor less than two, and the remaining 17.2\% have $s < 0.5 a$. 

Multiplicative factors slightly larger than one have been proposed in the literature to convert $s$ into $a$ \citep[e.g.][]{R10}. Different values of scaling factors are obtained depending on the assumed eccentricity distribution and on the metric used to compute the factor (median, mean $s/a$, mean $a/s$, mean $\log a/s$, etc.). The distribution of $s/a$ in Figure~\ref{fig:simsep} is almost symmetric, its mean is very close to the median. However, the distribution of the logarithm is skewed, and $\langle \log (s/a) \rangle = -0.073$ might suggest $a \approx 1.18 s$, while $\langle a/s \rangle = 1.59$. The correct approach is to de-convolve the observed distributions of $s$ from projection using the kernel from simulations (or its analytical approximation), instead of simple scaling. Example of such deconvolution is provided in the Appendix of \citep{MSC}.

\end{document}